\documentclass[12pt]{article}
\usepackage{graphicx}
\usepackage{amsmath,amssymb}
\newcommand{\dbox}{\,\raise2pt\hbox{\fbox{\rule{2.5pt}{0pt}\rule{0pt}{2.5pt}}}\,}
\newcommand{\qed}{\,\raise0pt\hbox{\mbox{\rule{6.5pt}{6.5pt}}}}
\newcommand{\bra}[1]{\mbox{$\langle #1 |$}}
\newcommand{\ket}[1]{\mbox{$| #1 \rangle$}}

\def\theequation{\thesection.\arabic{equation}}
\makeatletter
\@addtoreset{equation}{section}
\makeatother
\begin{document}
\setlength{\baselineskip}{7mm}

\begin{titlepage}
 \begin{normalsize}
  \begin{flushright}
        September 2012
 \end{flushright}
 \end{normalsize}
 \begin{LARGE}
   \vspace{1cm}
   \begin{center}
Gauge invariant and gauge fixed actions for various higher-spin fields 
from string field theory
\\
       \end{center}
 \end{LARGE}
  \vspace{5mm}
 \begin{center}
    Masako {\sc Asano} 
\\
      \vspace{4mm}
        {\sl Faculty of Science and Technology}\\
        {\sl Seikei University}\\
        {\sl Musashino-shi, Tokyo 180-8633, Japan}\\
      \vspace{1cm}

  ABSTRACT\par
 \end{center}
 \begin{quote}
\begin{normalsize}
We propose a systematic procedure for extracting gauge invariant and gauge fixed actions for various higher-spin gauge field theories 
from covariant bosonic open string field theory. 
By identifying minimal gauge invariant part for the original free string field theory action, we explicitly construct a class of covariantly gauge fixed actions with BRST and anti-BRST invariance.
By expanding the actions with respect to the level $N$ of string states,
the actions for various massive fields including higher-spin fields are systematically obtained.
As illustrating examples,
we explicitly investigate the level $N\le 3$ part and obtain the consistent actions for 
massive graviton field, massive 3rd rank symmetric tensor field, or antisymmetric field.
We also investigate the tensionless limit of the actions and 
explicitly derive the gauge invariant and gauge fixed actions for general rank $n$ symmetric and anti-symmetric tensor fields.
\end{normalsize}
 \end{quote}

\end{titlepage}
\vfil\eject

\section{Introduction}
Construction of various massive or massless higher spin gauge field theories have been attracted much interest\,\cite{Sorokinetal}.
It is not straightforward to construct a simple quadratic action for a given higher spin field with any spin or symmetry, not to mention the interaction part.
One of the several approaches to construct consistent higher spin field theories is the string field theory motivated approach\,\cite{Bengtsson:1986ys,Ouvry:1986dv,Francia:2002pt, Sagnotti:2003qa,Francia:2006hp,Sagnotti:2011qp} since
the theory contains infinite tower of massive fields provided by various types of higher rank tensor fields.
However, though the spectrum of the string theory is well-understood from the viewpoint of first quantized worldsheet conformal field theory, 
complete structure of each field in terms of quantum field theory 
is not clear even at the quadratic level.
Thus it is still not easy to extract simple quadratic actions for general higher spin fields from the string field theory.

For lower spin fields, the gauge invariant actions for vector field $A_\mu$ or linearized graviton field $h_{\mu\nu}$ and the anti-symmetric tensor field $B_{\mu\nu}$ can easily be extracted from the 
massless part of quadratic open or closed string field theory.
For higher spin fields, 
it is known that a class of massless gauge invariant action for a certain set of 
symmetric tensor fields of spin up to $s$, which is called the `triplets,' is derived from the tensionless ($\alpha'\!\rightarrow \!\infty$) limit of bosonic string theory\,\cite{Bengtsson:1986ys,Ouvry:1986dv,Francia:2002pt, Sagnotti:2003qa}.
For more general massive higher spin fields of mixed symmtery, however, it is rather technically involved to extract the corresponding simple and consistent actions from the string field theory
since many such fields are intertwined with one another even in the quadratic level.
One reason why these actions are complicated is that there exist a lot more degrees of freedom than required for constructing the consistent quadratic actions
in the original string field theory,
though those extra degrees of freedom are necessary for constructing consistent interacting string field theory.

All the information of string theory as a quantum field theory should in principle be included in the string field theory.
For example, it has become evident that 
the covariant open bosonic string field theory\,\cite{Siegel:1985tw,Banks:1985ff,Itoh:1985bb,Witten:1985cc} 
indeed contains the non-perturbative information as well as the perturbative one\,\cite{Sen:1999nx, Schnabl:2005gv}.
In particular, 
the covariant string field theory contains every fundamental properties of perturbative 
quantum field theory such as gauge invariance and the gauge fixing procedure from which propagators are derived. 
Up to the present, 
various types of gauges and the corresponding gauge fixed actions have been provided 
for bosonic string\,\cite{Schnabl:2005gv,Siegel:1984wx,Bochicchio:1986bd,Bochicchio:1986zj,Thorn:1986qj,Asano:2006hk,Asano:2008iu,Kiermaier:2007jg,Siegel:1984xd,Zwiebach:1992ie,Asano:2012sk} or superstring \,\cite{Preitschopf:fc,Arefeva:1989cp,Kroyter:2012niTorii:2012nj,Kroyter:2009rn,Kohriki:2011zzKohriki:2011pp,Kroyter:2009zi,Berkovits:2001im,Michishita:2004by} field theories.
Among such a variety of gauges, we could choose appropriate ones in accordance with the purpose of investigation. 

In this paper, we would like to give a systematic procedure for extracting gauge invariant and gauge fixed actions for various gauge field theories from covariant bosonic open string field theory by developing the discussion given in \cite{Asano:2006hk,Asano:2008iu}.
In ref.\cite{Asano:2006hk}, it is shown that the quadratic action for open string field theory\,\cite{Siegel:1985tw} is divided into two gauge invariant part: One is given by 
\begin{equation}
S_{\rm inv.}^{\rm min} =-\frac{1}{2} \left\langle  \phi^{(0)}, c_0L_0(1-P_0)\phi^{(0)}\right\rangle
\end{equation}
where $L_0$ and $1-P_0$ are zero mode of Virasoro algebra and a certain projection operator respectively, and the other is the action for auxiliary fields without kinetic terms.
The former action $S_{\rm inv.}^{\rm min}$ only consists of $b_0=0$ part of string fields $\phi^{(0)}$ and it explicitly gives minimal gauge invariant action for all the physical degrees of freedom in each level $N$. 
Here, $N$ is given by $L_0=\alpha' p^2 +N-1$ and the mass $m$ is determined by the on-shell condition $L_0=0$ as $m^2 = (N\!-\!1)/\alpha'$.
For example, for the massless level $N=1$, the action $S_{\rm inv.}^{\rm min}$ exactly gives
$-\frac{1}{4}F_{\mu\nu}F^{\mu\nu}$, which is the well-known action for the gauge field $A_\mu$. 
For the higher level massive fields, we can also extract from $S_{\rm inv.}^{\rm min}$ the minimal gauge invariant quadratic action without auxiliary fields.
Thus the result should be useful for constructing gauge invariant actions of various massive higher spin fields:
We can expect to obtain the simple gauge invariant actions for the corresponding fields without including redundant fields.

Furthermore, as for the gauge fixed action, we may expect that we can identify the minimal gauge fixed action part in the original gauge fixed action, 
which should correspond to the gauge fixed action for $S_{\rm inv.}^{\rm min}$. 
In fact, we shall explicitly construct such minimal gauge fixed action for the `$a$-gauge,' given in ref.\cite{Asano:2006hk}.
The $a$-gauge gives a one-parameter family of covariant gauges which include the Landau and the Feynman gauge which are well-known gauges for the massless vector gauge field.
Thus the result should provide a useful tool for perturbative analysis of the various higher spin fields contained in string theory.
Note that such systematic construction of gauge fixed actions for higher spin fields 
has not analyzed before. 
The gauge fixed action we obtain indeed has satisfactory properties for perturbative analysis:
It is invariant under BRST and anti-BRST transformations and the propagators are systematically constructed from it.
We also generalize the $a$-gauge to admit as many number of parameters as possible.

We explicitly see the structure of the minimal gauge invariant and gauge fixed action for lower level $N\le 3$.
In particular, we find that the gauge invariant action for massive level  $N=2$ exactly coincides with that for weak massive graviton field with Fierz-Pauli mass term\,\cite{Fierz:1939ix} in the St\"{u}ckelberg formalism (e.g., \cite{Hinterbichler:2011tt}).
For $N=3$, the action can be divided into the part which contains a rank 3 symmetric tensor field and the part which contains a rank 2 anti-symmetric tensor field.

We also discuss the tensionless limit of our minimal gauge invariant and gauge fixed actions by taking the leading terms in the $\alpha'\rightarrow \infty$ limit.
The result for gauge invariant action exactly coincides with the one given in\,\cite{Bengtsson:1986ys,Ouvry:1986dv}.
We explicitly construct the gauge fixed action for the `triplets,'
which has the simple form similar to that for the vector field.  

\bigskip

The organization of the paper is as follows. 
In the next section~2, we briefly review the structure of $S_{\rm inv.}^{\rm min}$ 
and the $a$-gauge given in ref.\cite{Asano:2006hk}.
After extending the $a$-gauge to admit multiple number of parameters, we construct the corresponding gauge fixed action $S^{(2)}_{{\rm GF}, \{a\}}$.
Then in section~3, by performing suitable redefinitions of string fields, we extract the minimal gauge fixed action $S^{\rm min}_{\{\alpha\}}$ from $S^{(2)}_{{\rm GF}, \{a\}}$.
After that, we identify BRST and anti-BRST transformations for $S^{\rm min}_{\{\alpha\}}$ and explicitly construct the propagators.
In section~4, we explicitly calculate the gauge invariant action 
$S^{\rm min}_{\rm inv.}$ for the level  $N\le 3$ and the gauge fixed action $S^{\rm min}_{\{\alpha\}}$ for $N=1,2$ and investigate the structure of them.
We also discuss the massless $m\rightarrow 0$ limit of the actions.
In section~5, we take the tensionless limit of our minimal gauge invariant and gauge fixed actions and explicitly construct the actions for general totally symmetric and anti-symmetric tensors.
In the last section~6, we summarize the results and give some discussions.
In Appendix A and B, we give useful formulas for string Fock space and the inner product necessary for understanding the gauge fixing and the actions.
In Appendix C, we construct another anti-BRST transformation which commutes with the BRST transformation given in section~3.

\section{Generalized $a$-gauges for open string field theory}
The quadratic action for covariant bosonic open string field  theory is given by~\cite{Siegel:1985tw,Banks:1985ff,Itoh:1985bb} 
\begin{equation}
S^{(2)} = -\frac{1}{2} \langle \Phi_1, Q \Phi_1  \rangle.
\label{eq:Squad}
\end{equation}
Here, $Q$ is the BRST operator%
\footnote{Note that this is the Noether charge for worldsheet BRST transformation.
Do not confuse with the BRST transformation for string fields which appear in later sections.}
\cite{Kato:1983im} and
$\Phi_1$ is the Grassmann-odd string field which is expanded by open string Fock states $\ket{f_i}_1$ of ghost number 1 associated with the corresponding fields $\phi_{f_i,1}$ as $\Phi_1 = \sum_i \ket{f_i}_1\, \phi_{f_i,1}.$
The action $S^{(2)}$ is invariant under the gauge transformation
\begin{equation}
\delta\Phi_1 = Q \Lambda_0 
\label{eq:gaugetr1}
\end{equation}
where $\Lambda_0$ is the Grassmann-even string field of ghost number $0$.
See Appendix~A and B for detail of the action, inner product and the Fock space.%
\footnote{Also, see, e.g., ref.\cite{Taylor:2003gn} for generalities of string field theory.}

As we have shown in ref.\cite{Asano:2006hk}, the action $S^{(2)}$ is divided into two gauge invariant parts as $S^{(2)} = S_{\rm inv.}^{\rm min}+S_{\rm auxiliary}$ where
\begin{eqnarray}
S_{\rm inv.}^{\rm min} &=&-\frac{1}{2} \left\langle  \phi^{(0)}, c_0L_0(1-P_0)\phi^{(0)}\right\rangle,
\label{eq:Smininv}
\\
S_{\rm auxiliary} &=&
\frac{1}{2}\left\langle (\omega^{(-1)}+W_1\tilde{Q}\phi^{(0)}),c_0 M (\omega^{(-1)}+W_1\tilde{Q}\phi^{(0)}) \right\rangle.
\end{eqnarray}
Here, we have decomposed the string field $\Phi_1$ and the BRST operator $Q$ with respect to the ghost zero modes as $\Phi_1=\phi^{(0)}+c_0 \omega^{(-1)}$ and $Q = \tilde{Q} + c_0 L_0 + b_0 M$.
Note that $\phi^{(0)}$ and $\omega^{(-1)}$ belong to the space ${\cal F}^0$ and ${\cal F}^{-1}$ defined by (\ref{eq:A1}) and $b_0 \phi^{(0)} = b_0 \omega^{(-1)}=0$.
The operator $W_n$ which is defined on the space ${\cal F}^n$ is given by 
\begin{equation}
W_n =\sum_{i=0}^{\infty}\, (-1)^i \frac{(n+i-1)!}{[(n+i)!]^2\, i!\,(n-1)!}\,
(M^-)^{n+i} M^i 
\end{equation}
where 
\begin{equation}
M=-2 \sum_{n>0} n c_{-n} c_{n},\qquad
M^-= - \sum_{n>0} \frac{1}{2n}  b_{-n} b_{n} .
\label{eq:MM-}
\end{equation}
The projection operator $P_0$, which is defined on ${\cal F}^0$,
 is explicitly defined by $P_0=-\frac{1}{L_0} \tilde{Q}W_1\tilde{Q}$ and it satisfies the relation $(1-P_0)\tilde{Q}=\tilde{Q}(1-P_0)=0$.
From the properties of $P_0$, we see that the action $S_{\rm inv.}^{\rm min}$ is invariant under the transformation 
\begin{equation}
\delta \phi^{(0)} = \tilde{Q} \lambda^{(-1)}
\label{eq:tildeQtr}
\end{equation}
where $\lambda^{(-1)}$ is the $b_0=0$ part of $\Lambda_{0}$: $\lambda^{(-1)}=b_0c_0\Lambda_{0}$.
Also, the equations of motion for $S_{\rm inv.}^{\rm min}$ is given by
\begin{equation}
L_0(1-P_0) \phi^{(0)}\,\left(
=(L_0+\tilde{Q}W_1\tilde{Q}) \phi^{(0)}
\right)=0.
\end{equation}
The operators $M$ and $M^-$, which are given by (\ref{eq:MM-}), with $M_z\equiv \frac{1}{2} \sum_{n>0} (c_{-n} b_n -b_{-n} c_n)$ constitute the SU(1,1) algebra.
Note that $2M_z$ counts the ghost number of non-zero mode part for $b_n$ and $c_n$.
All states in the space ${\cal F}^{n}$ are classified by  
the SU(1,1)-spin $s$ with $s\in \{ \frac{|n|}{2} + \pmb{Z}_{\ge 0} \}$ where $\pmb{Z}_{\ge 0}$ is the set of non-negative integers\,\cite{Asano:2006hk}.
Since $(1-P_0) \ket{f}=0$ for an $SU(1,1)$-spin $>0$ state $\ket{f}\in {\cal F}^0$, we can show that only the SU(1,1)-spin $=0$ string fields,
which can be specified by the condition 
\begin{equation}
M\phi^{(0)}=0
\end{equation}
on $\phi^{(0)}$, appear in the action $S_{\rm inv.}^{\rm min}$.

On the other hand, $S_{\rm auxiliary}$ does not contain $L_0$ and is indeed the action for only auxiliary fields. 
Thus we can consistently decouple $S_{\rm auxiliary}$ from $S_{\rm inv.}^{\rm min}$ and 
eliminate the degrees of freedom corresponding to $\omega^{(-1)}$ when we investigate the properties concerning the gauge invariance or gauge fixing of the quadratic part of the action.%
\footnote{Note that if our aim is to investigate the quantum properties of strings including interactions, we cannot ignore $S_{\rm auxiliary}$ since the $\omega^{(-1)}$-fields couple to other fields through the interaction terms.}

Returning to the original quadratic action $S^{(2)}$, we briefly review the gauge fixing condition ($a$-gauge) proposed in refs.\cite{Asano:2006hk, Asano:2008iu} and after that we generalize the condition to admit multiple parameters.
The $a$-gauge fixing condition is defined by imposing on $\Phi_1(=\phi^{(0)}+c_0\omega^{(-1)})$ the condition 
\begin{equation}
\frac{1}{1-a}(b_0 +ab_0c_0 W_{1}\tilde{Q}) 
\Phi_1\left(= {\rm bpz}({\cal O}_a^{\langle 3 \rangle})  \Phi_1\right) =0 
\label{eq:gaugephi1}
\end{equation}
where $a$ is a real parameter satisfying $a\ne 1$ including $|a|=\infty$.
Then the gauge fixed action is given by
\begin{equation}
S^{(2)}_{{\rm GF},a} = 
-\frac{1}{2}  \sum_{n=-\infty}^{\infty} \left\langle \Phi_n, Q  \Phi_{-n+2} \right\rangle
+  \sum_{n=-\infty}^{\infty} 
\left\langle {\cal O}_a^{\langle -n+4 \rangle}  {\cal B}_{-n+4} , \Phi_{n}  \right\rangle 
\label{eq:SGF}
\end{equation}
where $\Phi_n$ and ${\cal B}_{n}$ are Grassmann odd string fields of ghost number $n$.
The operator ${\cal O}_a^{\langle n \rangle}$ is defined by
\begin{eqnarray}
{\rm bpz}({\cal O}_a^{\langle n+1 \rangle} ) &=& \frac{1}{1-a}(b_0 +ab_0c_0 W_{n-1}M^{n-2}\tilde{Q}) \hspace*{2.6cm} \mbox{for } n\ge 2,
\\
{\rm bpz}({\cal O}_a^{\langle -n+4 \rangle} ) &=& b_0 (1-P_{n-2})
\nonumber\\
&&\qquad+
\frac{1}{1-a}(b_0 P_{n-2} +ab_0c_0 \tilde{Q} M^{n-2} W_{n-1})\qquad \mbox{for }n\ge 2
\end{eqnarray}
where 
${\rm bpz}({\cal O}_a^{\langle n \rangle})$ is the BPZ conjugation of ${\cal O}_a^{\langle n \rangle}$ and operates on ghost number $-n+4$ string fields $\Phi_{-n+4}$.
The operator $P_n$ is the projection operator defined for a state $\ket{f^{(n)}} \in {\cal F}^n$ with $L_0\ne 0$ as given explicitly in Appendix~A.
Note that as long as $|a|<\infty$, the operator ${\rm bpz}({\cal O}_a^{\langle n \rangle} )$ can be replaced by the following simpler form 
\begin{equation}
{\rm bpz}({\cal O}_a^{\langle n+1 \rangle} ) =b_0 +ab_0c_0 W_{n-1}M^{n-2}\tilde{Q} ,
\quad
{\rm bpz}({\cal O}_a^{\langle -n+4 \rangle} )= b_0 
 +ab_0c_0 \tilde{Q} M^{n-2} W_{n-1}
\label{eq:conda+-}
\end{equation}
for $n\ge 2$.
For convenience of extracting the minimal gauge fixed action from the original action 
(\ref{eq:SGF}), we use the second definition (\ref{eq:conda+-}).
In fact, the gauge-fixed action for $|a|\rightarrow \infty$ can be consistently obtained by taking the limit of the action for finite $a$.

Furthermore, we can generalize the gauge fixing condition by replacing ${\cal O}_a$ with ${\cal O}_{\{a\}}$ as
\begin{eqnarray}
\!\!{\rm bpz}({\cal O}_{\{a\}}^{\langle n+1 \rangle} ) \!&=&\!\!\!\!\! 
\sum_{k\in \{\frac{n-1}{2}+\pmb{Z}_{\ge 0} \}} \!\!\!\!S_k
\!\!\left[
\!\frac{1}{1\!-\!a_{-n+1}^k }\!\!\left(b_0 +a_{-n+1}^kb_0c_0 W_{n-1}M^{n-2}\tilde{Q} 
\right)\right]
\; \mbox{for } n\!\ge\! 2,
\\
{\rm bpz}({\cal O}_{\{a\}}^{\langle -n+4 \rangle} ) \!&=& \!\!
\sum_{k\in \{\frac{n-2}{2}+\pmb{Z}_{\ge 0}\}} \!\!S_k
\left[
b_0 P_{n-2} +a^k_{n-2}b_0c_0 \tilde{Q} M^{n-2} W_{n-1}
\right]
\quad \mbox{for }n\ge 2\;\;
\end{eqnarray}
where 
each $a^k_n\,(\ne 1)$ is an independent parameter and
$S_k$ is the projection operator onto the space of the SU(1,1)-spin $k$ states. 
The explicit form of $S_k$ is given in Appendix~A.
We can easily show that the gauge fixing condition given by these generalized operators ${\cal O}_{\{a\}}$ give the consistent gauge fixed action $S^{(2)}_{{\rm GF},\{a\}}$ which is obtained by replacing ${\cal O}_a$ with ${\cal O}_{\{a\}}$ in $S^{(2)}_{{\rm GF},a}$.

In ref.\cite{Asano:2006hk}, we have shown that the action (\ref{eq:SGF}) for level $N=1$ part reduces to the well-known action for massless vector field $A_\mu$ plus  decoupled auxiliary field terms after certain field redefinitions:
\begin{eqnarray}
S_{a,\,N=1} &=&\int{d^{D}x}\left[
-\frac{1}{4}F_{\mu\nu}F^{\mu\nu}+B\partial_{\mu}A^{\mu}+\frac{\alpha}{2}B^2
+i\bar{\gamma}\,\partial_{\mu}\partial^{\mu}\gamma
\right.
\nonumber\\
&&\hspace{35mm}\left.-\frac{1}{2}\tilde{\chi}^2
+\frac{1}{2}\tilde\beta_{u_{\mu}}\tilde{u}^{\mu}+\frac{1}{4}\beta_v v
\right].
\label{eq:SaN1}
\end{eqnarray}
Here, $F_{\mu\nu}=\partial_\mu A_\nu-\partial_\nu A_\mu$ 
and $\alpha=(1-a)^{-2}$.
The first line of this action,
which includes ghost and anti-ghost fields ($\gamma$ and $\bar{\gamma}$),
and Nakanishi-Lautrup field $B$, has exactly the same form as the known covariantly gauge-fixed action for the original gauge invariant action
\begin{equation}
  S = \int{d^{D} x} \left[-\frac{1}{4}F_{\mu\nu}F^{\mu\nu}\right].
\label{eq:N1inv}
\end{equation}
In particular, $\alpha=1$ ($a=0$) and $\alpha=0$ ($a=\infty$) respectively correspond to Feynman and Landau gauges.
On the other hand, the three terms in the second line in (\ref{eq:SaN1}) are auxiliary terms 
completely decoupled from other fields.
In fact, as we will explicitly see in the next section, the gauge invariant action (\ref{eq:N1inv}) is nothing but the level $N=1$ part of the minimal gauge invariant action $S^{\rm min}_{\rm inv.}$ given in (\ref{eq:Smininv}).
Note that since $\phi^{(0)}$ for $N=1$ (and $N=2$) only contains SU(1,1)-spin $=0$ states, the generalization ${\cal O}_a \rightarrow {\cal O}_{\{a\}}$ given above does not affect the results.

\section{Minimal gauge fixed action for open string fields}
In this section, by performing suitable redefinitions of string fields, we will show that the gauge fixed action given in the previous section 
can be consistently divided into three parts 
$ S^{(2)}_{{\rm GF},\{a\}} = S_{\rm inv.}^{\rm min}+
S_{{\rm gh+gf},\{\alpha\}} + S'_{\rm auxiliary}$
where 
the first two parts are regarded as the gauge fixed action 
for the minimal gauge invariant action $S_{\rm inv.}^{\rm min}$, and the third part is for the auxiliary fields decoupling from the other two parts. 
We will show that the action $S^{\rm min}_{{\rm GF},\{a\}} =S_{\rm inv.}^{\rm min}+
S_{{\rm gh+gf},\{\alpha\}}$ without $S'_{\rm auxiliary}$ is 
indeed a consistent gauge fixed action which is 
invariant under BRST and anti-BRST transformations in itself,
and from which the propagator is explicitly calculated.

\subsection{Minimal gauge invariant and gauge fixed actions}
We divide the string fields $\Phi_n$ and ${\cal B}_n$ by ghost zero-modes as 
\begin{equation}
\Phi_n=\phi^{(n-1)} +c_0 \omega^{(n-2)},
\qquad 
{\cal B}_n = c_0 \beta^{(n-2)}.
\label{eq:phiomega}
\end{equation}
Here $\phi^{(n)}$, $\omega^{(n)}$ and $\beta^{(n)}$ are the string fields consist of states in the space ${\cal F}^n$. 
Note that here we drop $\beta^{(n-1)}$ part in ${\cal B}_n$ which does not appear in the action $S^{(2)}_{\rm GF},\{a\}$ because of the property ${\cal O}^{\langle n\rangle}_{\{a\}} ={\cal O}^{\langle n\rangle}_{\{a\}}c_0b_0 $.
We define the new string fields $\phi^{'(n)}$, $\omega^{'(n)}$ and $\beta^{'(n)}$ 
from $\phi^{(n)}$, $\omega^{(n)}$ and $\beta^{(n)}$
as
\begin{eqnarray}
\phi^{'(n)}&=& 
\sum_{k\in \{\frac{n-1}{2}+\pmb{Z}_{\ge 0} \}} 
(1- a^k_{n-1})
\tilde{Q} M^{n} W_{n+1} S_k[\tilde{Q}\phi^{(n)}] 
\qquad(n\ge 1),
\\
\phi^{'(-n)}&=& \phi^{(-n)}   \qquad(n\ge 0),
\\
\omega^{'(n)}&=&\omega^{(n)} + M^{n+1}W_{n+2} \tilde{Q}\phi^{(n+1)}
-\!\!\!
 \sum_{k\in \{\frac{n+2}{2}+\pmb{Z}_{\ge 0} \}} \!\!
\frac{S_k\left[ M^{n+1}W_{n+2} \beta^{(n+2)}\right] }{1-a^k_{-n-2}} 
\nonumber\\
&&
\qquad \qquad\qquad
+\,a^{n/2}_{n} S_{n/2}[\tilde{Q} M^nW_{n+1} \phi^{(n+1)}]
\qquad \quad (n\ge 0),
\\
\omega^{'(-1)} &=& \omega^{(-1)} + W_{1} \tilde{Q}\phi^{(0)}
- 
 \sum_{k\in \{\frac{1}{2}+\pmb{Z}_{\ge 0} \}} \!\!
 \frac{S_k \left[ W_{1}\beta^{(n+2)}\right] }{1-a^k_{-1}} ,
\\
\omega^{'(-n)}&=&\omega^{(-n)} + 
W_{n}M^{n-1} 
\tilde{Q}\phi^{(-n+1)}- W_{n}M^{n-1} \beta^{(-n+2)}
\quad (n\ge 2),
\end{eqnarray}
and 
\begin{equation}
\beta^{'(n)} = \beta^{(n)}
\quad (n\ge 1),
\qquad
\beta^{'(-n)} = \beta^{(-n)} -S_{n/2}[\tilde{Q}\phi^{-n-1} ]
\quad (n\ge 0).
\end{equation}
By using these new string fields, the action 
$ S^{(2)}_{{\rm GF},\{a\}} $ can be 
rewritten as the sum of three independent terms 
\begin{equation}
 S^{(2)}_{{\rm GF},\{a\}} = S^{\rm min}_{\rm inv.}+
S_{{\rm gh+gf},\{\alpha\}} + S'_{\rm auxiliary}
\end{equation}
where $S^{\rm min}_{\rm inv.}$ is given by (\ref{eq:Smininv}),
$S_{{\rm gh+gf},\{\alpha\}}$ is for $\phi^{'(n)}$ ($n \ne 0$) and 
$\beta^{'(n)}$,
and $S'_{\rm auxiliary}$ is for $\omega^{'(n)}$. 
Explicitly, $S_{{\rm gh+gf},\{\alpha\}} $ and $S'_{\rm auxiliary}$ are given by
\begin{eqnarray}
S_{{\rm gh+gf},\{\alpha\}} &=& 
-\sum_{n=1}^\infty \langle\phi^{'(n)}, c_0L_0(1-P_{-n})\phi^{(-n)}\rangle
\nonumber\\
&&
+
\sum_{n=0}^\infty \langle c_0 \beta^{'(n+1)}, W_{n+1}M^n\tilde{Q}\phi^{(-n)}\rangle
+\sum_{n=1}^\infty \langle c_0 \beta^{'(-n+1)}, M^n W_{n+1}\tilde{Q}\phi^{'(n)}\rangle
\nonumber\\
&&
+\sum_{n=0}^\infty 
\,\,\sum_{k\in \{\frac{n+1}{2} +\pmb{Z}_{\ge 0}\}} \!\!\alpha^k_{(-n+1,n+1)} \langle S_k\beta^{'(-n+1)}, c_0 M^n W_{n+1} 
S_k\beta^{'(n+1)}\rangle,
\label{eq:SFPgf}
\end{eqnarray}
and 
\begin{equation}
S'_{\rm auxiliary} = 
-\frac{1}{2} \sum_{n=-\infty}^\infty \langle M \omega^{'(n)}, c_0 \omega^{'(-n-2)}  \rangle
- \sum_{n=0}^\infty \langle S_{n/2} \omega^{'(n)}, c_0 \beta^{'(-n)}  \rangle . 
\end{equation}
Here we have used the new representations of the parameters given by
\begin{equation}
\alpha^k_{(1,1)}=\frac{1}{(1-a^k_{-1})^2},\qquad
\alpha^k_{(-n+1,n+1)} =\frac{1}{1-a^k_{-n-1}}\quad(n\ge 1) .
\end{equation}
For future convenience, we also define
\begin{equation}
\alpha^k_{(n+1,-n+1)} = \alpha^k_{(-n+1,n+1)}  \quad(n\ge 1) .
\end{equation}

In summary, the combination 
\begin{equation}
 S_{\{\alpha\}}^{\rm min} = S^{\rm min}_{\rm inv.} +S_{{\rm gh+gf},\{\alpha\}}
\end{equation} 
is consistently regarded as a minimal gauge fixed action for the gauge invariant action 
$S_{\rm inv.}^{\rm min}$ without $S'_{\rm auxiliary}$ part. 
This action $S_{\{\alpha\}}^{\rm min}$ only contains $\phi^{(n)}\in {\cal F}^n$ and $\beta^{(n)}\in \tilde{\cal F}^n$ string fields
whereas the original action $ S^{(2)}_{{\rm GF},\{a\}}$ contains all $\phi^{(n)}$, 
$\omega^{(n)}$ and $\beta^{(n)}$.
Here $\tilde{\cal F}^n$ is given by 
\begin{equation}
\tilde{\cal F}^n = {\cal F}^n \quad (n\ge 1),
\qquad
\tilde{\cal F}^{-n} = (1-S_{n/2}){\cal F}^{-n} \quad (n\ge 0).
\end{equation}
In the next two subsections, we will investigate the consistency of the minimal gauge fixed action $S^{\rm min}_{\{\alpha\}}$ by identifying the 
BRST and anti-BRST invariance and deriving the general form of 
the propagators for fields contained in the action $S_{\{\alpha\}}^{\rm min}$.

\subsection{BRST and anti-BRST symmetry}\label{sec:BaB} 
We first rewrite the gauge fixed action $S_{\{\alpha\}}^{\rm min}$, which is given by the sum of (\ref{eq:Smininv}) and (\ref{eq:SFPgf}), into the form
\begin{eqnarray}
S_{\{\alpha\}}^{\rm min} &=& 
-\frac{1}{2}\sum_{n=-\infty}^\infty \langle\phi^{(n+1)}, c_0L_0(1-P_{-n-1})\phi^{(-n-1)}\rangle
+
\sum_{n=-\infty}^\infty \langle c_0 \beta^{(n+2)}, \tilde{W}^{\langle -n \rangle}_{-} \tilde{Q}\phi^{(-n-1)} \rangle
\nonumber\\
&&
+\frac{1}{2}\sum_{n=-\infty}^\infty 
\,\,\sum_{k\in \{\frac{{\rm max}(|-n|,|n+2|) }{2} +\pmb{Z}_{\ge 0}\}} \!\!\alpha^k_{(-n,n+2)} 
\langle S_k \beta^{(-n)}, c_0 \tilde{W}_-^{\langle n+2 \rangle} S_k\beta^{(n+2)}\rangle,
\label{eq:Sminord}
\end{eqnarray}
or in the matrix form as
\begin{eqnarray}
\hspace*{-1cm}
S_{\{\alpha\}}^{\rm min}
&=&
-\frac{1}{2}  \sum_{n=-\infty}^{\infty} 
\nonumber\\
&&\hspace*{-1.5cm}
\Bigg\langle
(\phi^{(n+1)} \;\; \beta^{(n+2)} ), 
\left(
\begin{array}{cc}
   c_0L_0(1-P_{-n-1}) & -c_0 \tilde{Q}\tilde{W}^{\langle -n \rangle}_-  \\
    c_0 \tilde{W}^{\langle -n \rangle}_-\tilde{Q}     & -c_0 {\displaystyle \sum_k} \alpha^{k}_{-n, n+2}S^k \tilde{W}^{\langle -n  \rangle}_- 
  \end{array}
\right)
\left(
\begin{array}{c}
   \phi^{(-n-1)} \\
   \beta^{(-n)}
  \end{array}
\right)
\Bigg\rangle.
\label{eq:Sminmat}
\end{eqnarray}
Here and in the following, we omit the $'$ signs on  $\phi^{(n)}$ and $\beta^{(n)}$. 
We have also used the new operator    
$\tilde{W}^{\langle n \rangle}_-$ defined on ${\cal F}^n$ as
\begin{equation}
\tilde{W}^{\langle n\rangle}_-  = M^{n-1} W_n  
 \quad (n \ge 1),
\qquad
\tilde{W}^{\langle -n \rangle}_- =
W_{n+2} M^{n+1}  \quad (n \ge 0).
\label{eq:deftildeW}
\end{equation}

The structure of the action $S_{\{\alpha\}}^{\rm min}$ is as follows:
The SU(1,1)-spin $=s$ $(>0)$ part of each $\phi^{(n)}$ for $n=-2s,-2s+1,\cdots,2s$ provides the $2s$-th (anti-)ghost fields, and 
with these fields, the SU(1,1)-spin $=s$ $(>0)$ part of 
$\beta^{(n)}\in\tilde{\cal F}^n$ are needed to fix the gauge invariance.
For example, we first need SU(1,1)-spin $=\frac{1}{2}$ part of $\beta^{(1)}$ to 
fix the gauge invariance of the original gauge invariant action $S^{\rm min}_{\rm inv.}$, and   
also the SU(1,1)-spin $=\frac{1}{2}$ part of $\phi^{(1)}$ and $\phi^{(-1)}$ 
respectively for anti-ghost and ghost fields.
Then, we need SU(1,1)-spin $=1$ part of $\beta^{(2)}$ and $\beta^{(0)}$ for gauge fixing, and the same part of $\phi^{(\pm 2)}$ and $\phi^{(0)}$ 
for `(anti-)ghost for (anti-)ghost' (2nd rank (anti-)ghost) fields.

The gauge fixed action $S_{\{\alpha\}}^{\rm min}$ remains no gauge invariance, but 
is invariant under the following two types of transformations:
\begin{eqnarray}
\delta_{\rm B} \phi^{(n)} &=& \eta \beta^{(n)} \qquad (n\ge 1),
\label{eq:BRST1}
\\
\delta_{\rm B} \phi^{(-n)} &=& \eta \left[
S_{n/2 } \tilde{Q} \phi^{(-n-1)}+ 
M W_{n+2} M^{n+1} \beta^{(-n)} 
\right]
\qquad (n\ge 0),
\\
\delta_{\rm B} \beta^{(\pm n)} &=& 0 \qquad (n\ge 0)
\end{eqnarray}
and 
\begin{eqnarray}
\delta'_{\rm B} \phi^{(n)} &=& 
\eta' \left[
S_{n/2 }\tilde{Q} M^nW_{n+1}\phi^{(n+1)}+ 
M^{n+1} W_{n+2} \beta^{(n+2)} 
\right]
\qquad (n\ge 0),
\\
\delta'_{\rm B} \phi^{(-n)} &=& 
- \eta' W_{n} M^{n-1} \beta^{(-n+2)} 
\qquad (n\ge 1),
\\
\delta'_{\rm B} \beta^{(\pm n)} &=& 0 \qquad (n\ge 0)
\label{eq:antiBRST3}
\end{eqnarray}
where $\eta$ and $\eta'$ are Grassmann odd parameters.
We can regard these two transformations $\delta_{\rm B}$ and $\delta'_{\rm B}$ as the 
BRST and the anti-BRST transformations.
Indeed, these transformations have the nilpotency property:
\begin{equation}
\delta_{\rm B}{}^2 =0,\qquad \delta'_{\rm B}{}^2 =0.
\end{equation}
Note that $\delta_{\rm B}$ and $\delta'_{\rm B}$ do not commute with each other in general. 
In fact, we can deform $\delta_{\rm B}'$ (or $\delta_{\rm B}$) 
by including the transformations of the form $\delta_{\rm B}' \phi^{(-n)} \sim \beta^{(-n+1)} $ and $\delta_{\rm B}' \beta^{(n)} \sim \beta^{(n+1)}$
without destroying the nilpotency property. 
In Appendix~C, we explicitly show that we can find another anti-BRST transformation $\tilde{\delta}_{\rm B}'$ which commutes with $\delta_{\rm B}$:  $[\delta_{\rm B}, \delta'_{\rm B}]=0$.

\subsection{Propagator}
Now we explicitly calculate the propagator for the action $S_{\{\alpha\}}^{\rm min}$.
To find the inverse of the matrix in the action (\ref{eq:Sminmat}), 
we consider the following equation
\begin{equation}
\left(
\begin{array}{cc}
   \Delta^{\langle -n-1 \rangle}  & A^{\langle -n-2 \rangle}  \\
 \tilde{A}^{\langle -n-1 \rangle}   & B^{\langle -n-2 \rangle}
  \end{array}
\right)
\left(
\begin{array}{cc}
   c_0L_0(1-P_{-n-1}) & -c_0 \tilde{Q}\tilde{W}^{\langle -n \rangle}_-  \\
    c_0 \tilde{W}^{\langle -n \rangle}_-\tilde{Q}     & -c_0\sum_k \alpha^{k}_{-n, n+2}S^k \tilde{W}^{\langle -n \rangle}_- 
  \end{array}
\right)
=1 c_0
\end{equation}
defined on  $({\cal F}_{-n-1}\; \tilde{\cal F}_{-n} )^{\rm T}$.
This can be solved by using the relations 
\begin{equation} 
P_n = -\frac{1}{L_0}\tilde{Q} \tilde{W}_-^{\langle n+1 \rangle} \tilde{Q},
\qquad
\tilde{W}_-^{{\langle n+1 \rangle}} \tilde{Q}(1-P_{n}) {\cal F}^n =0
\end{equation}
and some other properties given in the Appendix~A.
The result is
\begin{eqnarray}
   \Delta^{\langle -n-1 \rangle} &=&
\frac{1}{L_0} \left[ 1-P_{-n-1} 
-\sum_k \alpha^k_{(-n, n+2)} \frac{1}{L_0} \tilde{Q} S_k \tilde{W}_-^{{\langle -n \rangle}}\tilde{Q}
\right],
\label{eq:propDelta}
\\
 A^{\langle -n-2 \rangle}  &=&
\frac{1}{L_0} \tilde{Q} \tilde{W}_-^{\langle -n \rangle}M,
\\
 \tilde{A}^{\langle -n-1 \rangle} &=& - \frac{1}{L_0} M  \tilde{W}_-^{\langle -n \rangle}\tilde{Q},
\\
B^{\langle -n-2 \rangle} &=& 0.
\end{eqnarray}
Note that 
\begin{equation}
{\rm bpz}( \Delta^{\langle n \rangle}   ) = \Delta^{\langle -n \rangle},
\end{equation}
and the propagator between $\phi^{(n)}$ and $\phi^{(-n)}$ is given by $\Delta^{\langle n \rangle}$ (or $\Delta^{\langle -n \rangle}$). 
This $\Delta^{\langle n \rangle}$ is the operator acting on string states.
To extract the usual $c$-number propagator between the two fields $A(p)$ and $A'(p')$, we first provide the string fields corresponding to each of the fields as 
\begin{equation}
\int \frac{d^D p}{(2\pi)^D}  \ket{f(p)}_A A(p)\qquad \mbox{and} \qquad 
\int \frac{d^D p'}{(2\pi)^D} \ket{f(p')}_{A'}A'(p').
\end{equation}
If $\ket{f_A(p)}$ and $\ket{f_{A'}(p')}$ belong to $\phi^{(n)}$ and $\phi^{(-n)}$ respectively, 
the $c$-number propagator $\Delta_{A(p),A'(p')}$ is given by calculating the 
inner product for the corresponding string states after removing the $\delta$-function for $p+p'$.
Explicitly, the result is given as follows:
\begin{equation}
\Delta_{A(p),A'(p')} \delta^D(p+p')=
\left\langle {\rm bpz}({f_{A'}(p')})\,| \,\Delta^{\langle n \rangle} c_0 {f_A(p)}
\right\rangle .
\label{eq:cprop}
\end{equation}

\section{Examples: gauge invariant and gauge-fixed actions for level $N \le 3$}
We have seen that the action $S_{\{\alpha\}}^{\rm min}$ 
given by (\ref{eq:Sminord}) or (\ref{eq:Sminmat}) 
is indeed the consistent gauge-fixed action for minimal gauge invariant action $S_{\rm inv.}^{\rm min}$ given by (\ref{eq:Smininv})
in the sense that it is invariant under BRST and anti-BRST transformations and from which the propagator can be consistently derived. 
In $S_{\{\alpha\}}^{\rm min}$, there are infinite tower of massive fields given by various string states included in $\phi^{(0)}$. 
We divide the action with respect to the level $N$ of string states as
\begin{equation}
S_{\{\alpha\}}^{\rm min} =\sum_{N=0}^{\infty} S_{\{\alpha\}, N}^{\rm min} 
=\sum_{N=0}^{\infty} \left(
S_{{\rm inv.}, N}^{\rm min} +  S_{{\rm gh+gf},\{\alpha\}}^N
\right).
\end{equation}
Then, $S_{\{\alpha\}, N}^{\rm min}$ represents the action for massive fields with $m^2= (N-1)/\alpha'$. 
In this section, we investigate the minimal gauge invariant action $S^{\rm min}_{{\rm inv.},N}$ for $N=1, 2, 3$ 
and gauge fixed action  $S_{\{\alpha\}, N}^{\rm min}$ for $N=1, 2$.
For each of these actions, we explicitly calculate the inner product of string states and 
obtain the action in the usual field theory form.
\subsection{Massless vector field $A_\mu$ ($N=1$)}
For $N=1$ and $N=2$, the only non-trivial string fields appearing in $S_{\{\alpha\}, N}^{\rm min}$ are $\phi^{(0)}$, $\phi^{(\pm 1)}$ and $\beta^{(1)}$.
Thus the action has the form 
\begin{eqnarray}
S_{\{\alpha\}, N\!=\!1,2}^{\rm min}
&=& 
-\frac{1}{2} \langle\phi^{(0)}, c_0L_0(1-P_{0})\phi^{(0)}\rangle
\nonumber\\
&&\hspace*{-1cm}
- \langle\phi^{(1)}, c_0L_0(1-P_{-1})\phi^{(-1)}\rangle
+ \langle c_0 \beta^{(1)}, c_0 W_{1} \tilde{Q}\phi^{(0)}\rangle
+\frac{1}{2}\alpha 
\langle \beta^{(1)}, c_0 {W}_{1} \beta^{(1)}\rangle
\label{eq:SminN12}
\end{eqnarray}
where the first line is the gauge invariant action part.
Note that 
we have only one parameter $\alpha(= \alpha^{1/2}_{(1,1)})$
since $\phi^{(0)}$ and $\beta^{(1)}$ have SU(1,1)-spin $=0$ and $1/2$ respectively.

For $N=1$, the string field $\phi^{(0)}$ contains only the massless vector field $A_\mu$:
\begin{equation}
\phi^{(0)}_{N=1} = \int \frac{d^D p}{(2\pi)^D} \frac{1}{\sqrt{\alpha'}} \alpha_{-1}^{\mu}
\ket{0,p;\downarrow} A_{\mu}(p)
\label{eq:phi0N1}
\end{equation}
and the gauge invariant action $S_{{\rm inv.},N=1}^{\rm min}$ is obtained after calculating the inner product as 
\begin{equation}
S_{{\rm inv.},N=1}^{\rm min} =
\int \frac{d^D p}{(2\pi)^D} \left[
-\frac{1}{2} A_{\mu}(-p) (p^2\eta^{\mu\nu} -p^\mu p^\nu) A_\nu(p)
\right],
\end{equation} 
which is written in the $x$-representation as 
\begin{equation}
S_{{\rm inv.},N=1}^{\rm min} =
\int {d^D x} \left[
-\frac{1}{4} F_{\mu\nu}F^{\mu\nu}\right].
\end{equation} 
The other fields in the gauge fixed action $S_{\{\alpha\}, N=1}^{\rm min}$ 
are given by
\begin{eqnarray}
\phi^{(1)}_{N=1} &=& \int \frac{d^D p}{(2\pi)^D} \frac{1}{\sqrt{\alpha'}} c_{-1}
\ket{0,p;\downarrow} i \bar{\gamma}(p)
\\
\phi^{(-1)}_{N=1} &=& \int \frac{d^D p}{(2\pi)^D} \frac{1}{\sqrt{\alpha'}} b_{-1}
\ket{0,p;\downarrow} {\gamma}(p)
\end{eqnarray}
and 
\begin{equation}
\beta^{(1)}_{N=1} = \int \frac{d^D p}{(2\pi)^D} \sqrt{2} c_{-1}
\ket{0,p;\downarrow} i {B}(p)
\end{equation}
where $B$ is Grassmann even field and $\gamma $ and $\bar{\gamma}$ are Grassmann odd fields.
The Grassmann parity of each field is determined so that the total Grassmann parity of $\phi^{(n)}$ and $\beta^{(n)}$ are odd and even respectively.
On the other hand, 
existence or non-existence of $i$ in the expansion is determined by Hermitian property of the corresponding string fields\,\cite{Asano:2006hk}:
For the field $\phi_f(p)$ associated with a state $\ket{f}$ in $\phi^{(n)}$ or in $\beta^{(n)}$, 
we assign $(\phi_f(p))^* = \epsilon_f \phi_f(-p)$ if ${\rm bpz}(\ket{f}) = \epsilon_f \bra{f}$.

By substituting these string fields into (\ref{eq:SminN12}) and calculating the inner product, we obtain the action 
\begin{eqnarray}
S_{\alpha, N=1}^{\rm min} &=& 
\int \frac{d^D p}{(2\pi)^D} \left[
-\frac{1}{2} A_{\mu}(-p) (p^2\eta^{\mu\nu} -p^\mu p^\nu) A_\nu(p)
+ip^\mu A_{\mu}(p) B (-p) 
\right.
\nonumber\\
&& \qquad\qquad
\left.
+ \frac{\alpha}{2}B (-p) B (p)
-
i\bar{\gamma} (-p) p^2 \gamma(p)
\right].
\end{eqnarray}
The action in $x$-representation is given by 
\begin{equation}
S_{\alpha, N=1}^{\rm min} = 
\int {d^D x} \left[
-\frac{1}{4} F_{\mu\nu}F^{\mu\nu}
+B \partial_\mu A^{\mu} 
+ \frac{\alpha}{2}B^2
+ i \bar{\gamma} \partial_\mu \partial^\mu \gamma(p)
\right].
\label{eq:SminN1}
\end{equation}
This exactly coincides with the first line of (\ref{eq:SaN1}).
Note that this action is consistent not only for $D=26$, but for general dimension $D$.
In fact, for $N=1$, the action in the form (\ref{eq:SminN12}) is already consistent in general dimension $D$ since the relation $Q^2=0$ holds for all the level $N=1$ fields.

The BRST and the anti-BRST transformations given by (\ref{eq:BRST1})$\sim$(\ref{eq:antiBRST3}) are reduced to the well-known form as 
\begin{equation}
\delta_{\rm B} A_\mu = -\eta i \partial_\mu\gamma,
\qquad 
\delta_{\rm B} \bar{\gamma} =\eta B,
\qquad
\delta_{\rm B}B = \delta_{\rm B} \gamma =0
\label{eq:A1BRST}
\end{equation}
and
\begin{equation}
\delta_{\rm B}' A_\mu = \eta' \partial_\mu\bar{\gamma},
\qquad 
\delta_{\rm B}' {\gamma} = -i \eta' B,
\qquad
\delta_{\rm B}' B = \delta_{\rm B}' \bar{\gamma} =0
\label{eq:A1aBRST}
\end{equation}
where we have rescaled the Grassmann odd parameters as $\sqrt{2\alpha'}\eta\rightarrow \eta$ and $\sqrt{\alpha'/2}\eta'\rightarrow -\eta'$. 
In this level $N=1$, $[\delta_{\rm B},\delta_{\rm B}']=0$ holds. 

The propagator for $A_{\mu}$ and $A_{\nu}$ is calculated by substituting 
(\ref{eq:phi0N1}) into (\ref{eq:cprop}) and as a result we obtain the usual form 
for covariant gauge 
\begin{equation}
\Delta_{A_{\mu},A_{\nu}}=
\frac{1}{p^2}\left[
\eta^{\mu\nu} -(1-\alpha)\frac{p^\mu p^\nu}{p^2}
\right].
\end{equation}

\subsection{Massive graviton field $g_{\mu\nu}$ ($N=2$)}
Next, we consider the $N=2$ part of the gauge invariant and gauge fixed actions $S_{{\rm inv.}, N=2}^{\rm min}$ and $S_{\{\alpha\}, N=2}^{\rm min}$.
In this case, since $Q^2=0$ 
only holds for $D= 26$ as is the case for general higher level string states, 
we fix $D=26$ in order to obtain the consistent actions.
However, we will see that 
after we eliminate all string states by calculating the inner products, 
the action can be consistently extended to any spacetime dimension $D$.

In this level $N=2$, non-trivial string fields appearing in $S_{\{\alpha\}, N=2}^{\rm min}$ are given as follows:
\begin{eqnarray}
\phi^{(0)}_{N=2} \!\!&=& \!\!\int\! \frac{d^{26} p}{(2\pi)^{26}}
\left[
 \alpha_{-1}^{\mu}\alpha_{-1}^{\nu}\ket{0,p;\downarrow} h_{\mu\nu}(p)
 +i\alpha_{-2}^{\mu}\ket{0,p;\downarrow} g_{\mu}(p)
 +b_{-1}c_{-1}\ket{0,p;\downarrow} \phi(p)
\right]
\\
\phi^{(1)}_{N=2} \!\!&=&\!\! \int\! \frac{d^{26} p}{(2\pi)^{26}} 
\left[
\frac{i}{\sqrt{\alpha'}} \alpha^{\mu}_{-1}c_{-1}
\ket{0,p;\downarrow} \bar{\gamma}_\mu(p)
+ \frac{1}{\sqrt{\alpha'}}c_{-2}
\ket{0,p;\downarrow} \bar{\gamma}(p)
\right]
\label{eq:N2phi1}
\\
\phi^{(-1)}_{N=2} \!\!&=& \int\! \frac{d^{26} p}{(2\pi)^{26}} 
\left[
\frac{1}{\sqrt{\alpha'}}
\alpha^{\mu}_{-1} b_{-1}
\ket{0,p;\downarrow} {\gamma}_\mu(p)
+ \frac{i}{\sqrt{\alpha'}} b_{-2}
\ket{0,p;\downarrow} {\gamma}(p)
\right]
\label{eq:N2phim1}
\end{eqnarray}
and 
\begin{equation}
\beta^{(1)}_{N=2} = \int \frac{d^{26} p}{(2\pi)^{26}} 
\left[
i \alpha^{\mu}_{-1} c_{-1}
\ket{0,p;\downarrow} {B}_\mu(p)
+  \sqrt{2} c_{-2}
\ket{0,p;\downarrow} {B}(p)
\right].
\label{eq:N2beta1}
\end{equation}
Here, the symmetric tensor field $h_{\mu\nu} (=h_{(\mu\nu)})$, the vector field $g_{\mu}$ and the scalar field $\phi$ are Grassmann even fields that constitute the gauge invariant action. 
The remaining $B_\mu$ and $B$ are Grassmann even fields
which correspond to the generalized Nakanishi-Lautrup fields, and $\bar{\gamma}_\mu$, $\bar{\gamma}$, ${\gamma}_\mu$ and ${\gamma}$ are Grassmann odd fields.
Here $\bar{\gamma}_\mu$ and $\bar{\gamma}$ are anti-ghost fields, and
${\gamma}_\mu$ and ${\gamma}$ are ghost fields.
After eliminating string states by calculating inner products by substituting $\phi^{(0)}_{N=2}$ into (\ref{eq:SminN12}), we obtain 
the gauge invariant action as 
\begin{eqnarray}
S_{{\rm inv.},N=2}^{\rm min}&=&
\int \!\! \frac{d^{D} p}{(2\pi)^{D}} \left[
h_{\mu\nu}(-p) \left( (-\alpha' p^2-1)\eta^{\nu\sigma}  +2\alpha' p^{\nu} p^{\sigma}   \right) 
h^{\mu}{}_{\sigma}(p)
+\frac{1}{8} h(-p) h(p)
\right.
\nonumber\\
&&
+ \alpha' g_{\mu}(-p) (p^\mu p^\nu -p^2 \eta^{\mu\nu}  ) g_{\nu}(p)
+i \sqrt{\frac{\alpha'}{2}}  \left(4 h_{\mu\sigma} (-p) p^{\mu} +h(-p) p_{\sigma}     
\right) g^{\sigma} (p)
\nonumber\\
&&
+ \phi(-p) \left( \alpha' p^2 +\frac{13}{8} \right) \phi(p) 
+ \phi(-p) \left( 2 \alpha' p^\mu p^\nu  +\frac{3}{4} \eta^{\mu\nu} \right) h_{\mu\nu}(p) 
\nonumber\\
&&
\left.
+5i  \sqrt{\frac{\alpha'}{2}}
\phi(-p) p_\mu g^{\mu}(p)
\right]
\end{eqnarray}
where $h=h_\mu{}^\mu$.
This action is invariant under the gauge transformations derived from (\ref{eq:tildeQtr}):
\begin{eqnarray}
\delta h_{\mu\nu}(p) &=& i\sqrt{2 \alpha'} p_{(\mu}\lambda_{\nu)}(p)
+\frac{1}{2} \eta_{\mu\nu} \lambda(p),
\\
\delta g_{\mu}(p) &=& \lambda_{\mu}(p) -  i \sqrt{2 \alpha'} p_\mu  \lambda(p),
\\
\delta \phi(p) &=&  -  i \sqrt{2 \alpha'} p^\mu  \lambda_{\mu}(p) -  3 \lambda(p)
\end{eqnarray}
for arbitrary Grassmann even fields $\lambda_{\mu}(p)$ and $\lambda(p)$.
Here the parenthesis of indices denotes the symmetrization. 

In order to separate the two types of gauge transformations concerning $\lambda_{\mu}(p)$ and $\lambda(p)$ as much as possible and to make the action simpler, we define $h_{\mu\nu}'$, $g'_\mu$ and $\phi'$ 
from the original fields  as
\begin{equation}
h_{\mu\nu}' = \sqrt{2\alpha'} \left(h_{\mu\nu} -\frac{1}{20}(\phi +h)\eta_{\mu\nu}\right),
\qquad
g_\mu'=\sqrt{\alpha'} g_\mu
\qquad
\phi'=\frac{\sqrt{2\alpha'} }{10} (\phi+h) .
\end{equation}
Then the gauge invariant action and the gauge invariance in $x$-representation can be recast into the following simpler form 
\begin{eqnarray}
S_{{\rm inv.}}^{N=2}&=&
\int \!\! {d^D x}
\left[
2 [\sqrt{-g'}R']|_{\rm (2)} -\frac{m^2}{2} \left(  h^{'2} -h_{\mu\nu}' h^{'\mu\nu}\right)
\right.
\nonumber\\
&& \left. -\frac{1}{2} G_{\mu\nu}G^{\mu\nu} 
+2 m(h'_{\mu\nu} - h' \eta_{\mu\nu}) \partial^\mu g^\nu
- 2 \phi' 
\left( \partial_\mu\partial^\mu h' -  \partial_\mu\partial_\nu h^{'\mu\nu }  \right)
\right]
\label{eq:SinvN2f}
\end{eqnarray}
and 
\begin{eqnarray}
\delta h'_{\mu\nu} &=& 2 \partial_{(\mu}\lambda_{\nu)},
\label{eq:ginvN21}
\\
\delta \phi' &=& -m \lambda,
\\
\delta g'_{\mu} &=& m \lambda_{\mu} +  \partial_\mu  \lambda.
\label{eq:ginvN23}
\end{eqnarray}
Here, $m^2=1/\alpha'$, $h'=h'{}_\mu{}^\mu$, $G_{\mu\nu}$ is the field strength for $g'_{\mu}$:
\begin{equation}
G_{\mu\nu} = \partial_{\mu}g'_{\nu} -\partial_{\nu}g'_{\mu} ,
\end{equation}
 and 
\begin{equation}
[\sqrt{-g'}R']|_{\rm (2)} = \frac{1}{4}h'_{\mu\nu}\square
 h^{'\mu\nu}
-\frac{1}{2}h'_{\mu\nu} \partial^{\mu} \partial^{\rho} h^{'\nu}_{\rho}
-\frac{1}{4}h'
\square 
h' +\frac{1}{2}h' \partial^{\mu} \partial^{\nu} h'_{\mu\nu},
\end{equation}
which is the quadratic order of the Einstein action for the metric $g'_{\mu\nu} \sim \eta_{\mu\nu} +h_{\mu\nu}'$. 
We have also rescaled gauge parameters as $-\sqrt{2}\alpha' \lambda\rightarrow \lambda$ and 
$\alpha' \lambda_\mu \rightarrow \lambda_\mu$.
Note that the action (\ref{eq:SinvN2f}) and its gauge invariance can be applied for any spacetime dimension $D$, although the original action written by string fields are only consistent for  $D=26$.
The action includes the Einstein term for weak graviton field $h_{\mu\nu}$ plus mass terms. 
Also, there are two St\"{u}ckelberg-like fields $g'_{\mu}$ and $\phi'$ with which the gauge invariance is maintained after including the mass terms.
This means that such additional fields are already consistently included in the string field theory.
In fact, the action (\ref{eq:SinvN2f}) exactly coincides with that for the massive graviton field with Fierz-Pauli mass term\,\cite{Fierz:1939ix} in the St\"{u}ckelberg formalism\,\cite{Hinterbichler:2011tt}.

We can obtain the (anti-)ghosts and the gauge fixing terms for the gauge invariant action 
$S_{\rm inv.}^{N=2}$ 
by substituting $\phi^{(\pm 1)}_{N=2}$ and $\beta^{(1)}_{N=2}$ into the second line of (\ref{eq:SminN12}).
The result is 
\begin{eqnarray}
S_{{\rm gh+gf},\,\alpha}^{N=2} &=&
\beta^{\mu} \left[
\partial^\nu \left(h'_{\mu\nu}- \frac{1}{2} \eta_{\mu\nu} h' \right)
- \partial_\mu\phi' -m g_\mu
\right]
\nonumber\\
&&
-  \beta \left[ \partial^\mu g_\mu
+ m\phi' 
- \frac{m}{2} h'
\right]
+\frac{\alpha}{4} (\beta^2 +\beta_\mu\beta^\mu)
\nonumber\\
&& - i \bar{\gamma} (\partial_\nu \partial^\nu -m^2) \gamma
+i \bar{\gamma}_\mu (\partial_\nu \partial^\nu -m^2) \gamma^{\mu}
\label{eq:SFPgfN2}
\end{eqnarray}
where $\alpha$ is a parameter.
Then the total gauge fixed action is given by the sum 
$S_{\{\alpha\}}^{N=2}  = S_{\rm inv.}^{N=2}+  S_{{\rm gh+gf},\,\alpha}^{N=2}$.
Again, the BRST and the anti-BRST transformations for this action are calculated by
(\ref{eq:BRST1})$\sim$(\ref{eq:antiBRST3}) and the result is 
\begin{eqnarray}
&&\delta_{\rm B} h_{\mu\nu} = -2 \eta i \partial_{(\mu} \gamma_{\nu)},
\qquad 
\delta_{\rm B} g_{\mu} = - \eta i ( m \gamma_{\mu} + \partial_\mu \gamma) ,
\qquad 
\delta_{\rm B} \phi =  \eta  i m \gamma,
\nonumber\\
&& 
\delta_{\rm B} \bar{\gamma}_\mu =\eta B_\mu,
\qquad \qquad\quad\;
\delta_{\rm B} \bar{\gamma} =\eta B,
\end{eqnarray}
and
\begin{eqnarray}
&&\delta'_{\rm B} h_{\mu\nu} = \eta' i \partial_{(\mu} \bar{\gamma}_{\nu)},
\qquad 
\delta'_{\rm B} g_{\mu} =  \frac{i}{4}\eta'  (2 m \bar{\gamma}_{\mu} + \partial_\mu \bar{\gamma}) ,
\qquad
\delta'_{\rm B} \phi =  -\frac{i}{4} \eta'   m \bar{\gamma},
\nonumber\\
&&
\delta'_{\rm B} {\gamma}_\mu =\frac{1}{2}\eta' B_\mu,
\qquad\quad\;
\delta'_{\rm B} {\gamma} =\frac{1}{4}\eta' B
\end{eqnarray}
where we only present the fields which yield non-zero values.
As for the case of $N=1$, $[\delta_{\rm B},\delta_{\rm B}']=0$ also holds in this level $N=2$.

Next, we consider the massless limit of the action $S_{\{\alpha\}}^{N=2}  = S_{\rm inv}^{N=2}+  S_{{\rm gh+gf},\,\alpha}^{N=2}$.
In this limit, the gauge invariant action $S_{\rm inv.}^{N=2}$ becomes
\begin{equation}
S_{{\rm inv.}, m\rightarrow 0}^{N=2}
=
\int \!\! {d^D x}
\left[
2 [\sqrt{-g'}R']\Big|_{\rm 2} 
-\frac{1}{2} G_{\mu\nu}G^{\mu\nu} 
- 2 \phi' 
\left( \partial_\mu\partial^\mu h' -  \partial_\mu\partial_\nu h'{}^{\mu\nu }  \right)
\right]
\end{equation}
and the gauge transformations given by (\ref{eq:ginvN21})$\sim$(\ref{eq:ginvN23}) are reduced to
\begin{equation}
\delta h'_{\mu\nu} = 2 \partial_{(\mu}\lambda_{\nu)}
,\qquad
\delta g_{\mu} =   \partial_\mu  \lambda
,\qquad
\delta \phi' =0.
\label{eq:ginvN2}
\end{equation}
The scalar field $\phi'$ can be decoupled from other fields by redefining $h'_{\mu\nu}$ as
\begin{equation}
h'_{\mu\nu} \rightarrow h'_{\mu\nu} -\frac{2}{D-2}\phi'\eta_{\mu\nu}  .
\label{eq:hredefN2}
\end{equation} 
Then the action is rewritten as
\begin{equation}
S_{{\rm inv.}, m\rightarrow 0}^{N=2}
=
\int \!\! {d^D x}
\left[
2 [\sqrt{-g'}R']\Big|_{\rm 2} 
-\frac{1}{2} G_{\mu\nu}G^{\mu\nu} 
-\frac{2(D-1)}{D-2}\partial_\mu\phi' \partial^{\mu}\phi' 
\right]
\end{equation} 
which is the sum of the actions for 
(weak) graviton field, vector field and the scalar field.
The gauge transformations remain the same form as (\ref{eq:ginvN2}) after the redefinition of $h'_{\mu\nu}$.
The (anti-)ghosts and gauge fixing terms $S_{{\rm gh+gf},\,\alpha}^{N=2}$ in the $m\rightarrow 0$ limit is given after the redefinition of $h'_{\mu\nu}$ by
\begin{eqnarray}
S_{{\rm gh+gf},\,\alpha,m\rightarrow 0}^{N=2} &=&
\beta^{\mu} \left[
\partial^\nu \left(h'_{\mu\nu}- \frac{1}{2} \eta_{\mu\nu} h' \right)
\right]
+\frac{\alpha}{4} \beta_\mu\beta^\mu
+i \bar{\gamma}_\mu \partial_\nu \partial^\nu \gamma^{\mu}
\nonumber\\
&&
-  \beta \partial^\mu g_\mu
+\frac{\alpha}{4} \beta^2 - i \bar{\gamma} \partial_\nu \partial^\nu \gamma.
\end{eqnarray}
Here, the first line and the second line are (anti-)ghost and gauge fixing terms for $h'_{\mu\nu}$ and $g_\mu$ respectively.


\subsection{Gauge invariant action for massive symmetric tensor field $A_{\mu\nu\rho}$ and anti-symmetric tensor field $B_{\mu\nu}$ ($N=3$)}
For the next level $N=3$, we only consider the gauge invariant action.
The string field $\phi^{(0)}$ expanded by the $N=3$ string states is given as follows:
\begin{eqnarray}
\phi^{(0)}_{N=3} \!&=& \!\int\! \frac{d^{26} p}{(2\pi)^{26}}
\Big[
 \alpha_{-1}^{\mu}\alpha_{-1}^{\nu}\alpha_{-1}^{\rho}\ket{0,p;\downarrow} A_{\mu\nu\rho}(p) 
  +\alpha_{-3}^{\mu}\ket{0,p;\downarrow} D_{\mu}(p)
+b_{-1}c_{-1}  \alpha_{-1}^{\mu}\ket{0,p;\downarrow} C_{\mu}(p)
\nonumber\\
&& 
 +\, i( \alpha_{-1}^{\mu}\alpha_{-2}^{\nu}+\alpha_{-1}^{\nu}\alpha_{-2}^{\mu} )h_{(\mu\nu)}(p)
+i( \alpha_{-1}^{\mu}\alpha_{-2}^{\nu}-\alpha_{-1}^{\nu}\alpha_{-2}^{\mu} )B_{[\mu\nu]}(p)
\nonumber\\
&& 
 +\, i (2b_{-1}c_{-2} + b_{-2}c_{-1})\ket{0,p;\downarrow} \phi_{s=0}(p)
+ i (2b_{-1}c_{-2} - b_{-2}c_{-1})\ket{0,p;\downarrow} \phi_{s=1}(p)
\Big].
\label{eq:N3phi}
\end{eqnarray}
The minimal gauge invariant action $S_{\rm inv.}^{N=3}$ is calculated 
by substituting $\phi^{(0)}_{N=3}$ into the general action (\ref{eq:Smininv}).
Also, gauge transformation for each field is given by substituting the string field 
$\lambda^{(-1)}_{N=3}$ expanded by the corresponding string states into (\ref{eq:tildeQtr}).
Note that one of the scalar fields $\phi_{s=1}(p)$ given in (\ref{eq:N3phi}) does not appear in the gauge invariant action $S_{\rm inv.}^{N=3}$ since it is associated with the SU(1,1)-spin $=1$ string state. 
Thus, the action $S_{\rm inv.}^{N=3}$ consists of 
a 3rd rank symmetric tensor field $A_{\mu\nu\rho}$, symmetric and anti-symmetric 2nd rank tensor fields $h_{(\mu\nu)}$ and $B_{[\mu\nu]}$, two vector fields $D_\mu$ and $C_{\mu}$, and a scalar field $\phi_{s=0}$.

The resulting action and the gauge transformations in terms of the above original fields are complicated. 
However, it can be expressed in a slightly simpler form if we perform the following field redefinitions:
\begin{eqnarray}
A_{\mu\nu\rho}' &=& 
\sqrt{2\alpha'} \left[
A_{\mu\nu\rho} -\frac{1}{8}\eta_{(\mu\nu}A_{\rho)\sigma}{}^\sigma
-\frac{1}{24}\left(
\eta_{(\mu\nu}D_{\rho)} +\eta_{(\mu\nu}C_{\rho)}
\right)
\right],
\label{eq:N3Sredef1}
\\
h_{(\mu\nu)}' &=& \sqrt{2\alpha'} \left[h_{\mu\nu} -\frac{1}{18}(\phi_{s=0}+h)\eta_{\mu\nu}\right],
\\
D_{\mu}' &=& \frac{\sqrt{2\alpha'} }{24}(D_\mu +13 C_\mu +3 A_{\mu\nu}{}^\nu),
\\
\phi' &=& -\frac{\sqrt{2\alpha'} }{3}(h+4\phi_{s=0}),
\label{eq:N3Sredef4}
\\
B_{[\mu\nu]}' &=& 4 \sqrt{2\alpha'} B_{[\mu\nu]} ,
\label{eq:N3Sredef5}
\\
C_{\mu}' &=& \frac{2\sqrt{2\alpha'} }{8}(3 D_\mu - C_\mu - A_{\mu\nu}{}^\nu)
\label{eq:N3Sredef6}
\end{eqnarray}
where
$h=h_{\mu}{}^{\mu}$ 
and the summation is taken in $D=26$ dimensional spacetime indices.
In terms of these new fields with primes, the action is divided into two independent parts
\begin{equation}
S_{\rm inv.}^{{\rm min};\, N=3}
= S_{\rm inv.}^{N=3,{\rm S}}
+S_{\rm inv.}^{N=3,{\rm A}}.
\end{equation}
The first part $S_{\rm inv.}^{N=3,{\rm S}}$ consists of 3rd rank symmetric tensor field 
$A'_{\mu\nu\rho}$ with lower rank fields: $h'_{(\mu\nu)}$, $D'_\mu$, and $\phi'$.
The other part $S_{\rm inv.}^{N=3,{\rm A}}$ is the action for the anti-symmetric field
$B'_{\mu\nu}$ with vector field $C'_{\mu}$.
We will see that these two actions $S_{\rm inv.}^{N=3,{\rm S}}$ and $S_{\rm inv.}^{N=3,{\rm A}}$ 
are separately invariant under independent gauge transformations.
In the following, we will explicitly investigate these two actions and their gauge transformations separately. 
We also investigate the massless limit of each action and discuss the relation to the known action for higher-spin fields obtained e.g., from the tensionless limit of open strings\,\cite{Bengtsson:1986ys, Ouvry:1986dv, Francia:2002pt, Sagnotti:2003qa}.

\subsubsection{Action for massive symmetric tensor field $A_{\mu\nu\rho}$ with
$h_{\mu\nu}$, $D_\mu$ and $\phi$}
By using the redefined fields given in (\ref{eq:N3Sredef1})$\sim$(\ref{eq:N3Sredef4}), 
the action $S_{\rm inv.}^{N=3,{\rm S}}$ is expressed as
\begin{eqnarray}
S_{\rm inv.}^{N=3,{\rm S}} &=&
-\frac{3}{2} \int\!\!{d^{D}x} 
\bigg\{
 -\Big[ A_{\mu\nu\rho} \Box A^{\mu\nu\rho}
+3 \partial_\rho A_{\mu\nu}{}^{\rho}  \partial_\sigma A^{\mu\nu\sigma} 
+ 3 \partial^\mu A_{\mu\nu}{}^{\nu}  \partial^\sigma A_{\sigma\rho}{}^{\rho} 
+3 A_{\mu\nu}{}^{\nu} \Box A^{\mu}{}_{\rho}{}^{\rho} 
\nonumber\\
&&
\qquad \qquad 
-\, 4D_\mu \left(\partial^\mu \partial^\nu - \eta^{\mu\nu} \Box \right)  A_{\nu\rho}{}^{\rho}
- \,4 D^{\rho} \partial^\mu \partial^\nu  A_{\mu\nu\rho} 
+\, 2 \partial_\mu D^{\mu} \partial_\nu  D^{\nu} \Big] 
\nonumber\\
&&
\quad + \,
(m A_{\mu\nu\rho} -2\partial_{(\mu}h_{\nu\rho)}  )
(m A^{\mu\nu\rho} -2\partial^{(\mu}h^{\nu\rho)}  )
\nonumber\\
&& 
\quad - \, 3 (m A_{\mu\rho}{}^{\rho} -2\partial_{(\mu}h_{\nu\rho)}\eta^{\nu\rho}  )
(m A^{\mu}{}_{\sigma}{}^{\sigma} -2\partial^{(\mu}h^{\sigma\lambda)}\eta_{\sigma\lambda}  )
\nonumber\\
&& 
\quad + \, 4 \phi \left( m\partial^\mu D_\mu + 3m\partial^\mu A_{\mu\nu}{}^{\nu} 
-2 \Box h_\mu{}^{\mu} -2 \partial^\mu \partial^{\nu} h_{\mu\nu} 
\right)
-2m^2\phi^2 +8\phi\Box \phi\bigg\}
\label{eq:N3S}
\end{eqnarray}
where $m=\sqrt{2/\alpha'}$ and we have omitted the prime $'$ on each field
defined by (\ref{eq:N3Sredef1})$\sim$(\ref{eq:N3Sredef4}).
This action is invariant under the gauge transformations
\begin{eqnarray}
\delta A_{\mu\nu\rho} &=&\partial_{(\mu} \lambda_{\nu\rho)},
\label{eq:N3Str1}
\\
\delta h_{\mu\nu} &=& \frac{m}{2} \lambda_{\mu\nu} + 2 \partial_{(\mu} \lambda_{\nu)},
\\
\delta D_{\mu} &=& - \partial^\nu \lambda_{\mu\nu} + 2 m \lambda_{\mu},
\label{eq:N3gaugeD}
\\
\delta \phi &=& 2 \partial_\mu \lambda^{\mu}
\label{eq:N3Str4}
\end{eqnarray}
where $\lambda_{\mu\nu} $ and $\lambda_{\mu}$ are symmetric tensor field and vector field respectively.
Note that this action in the form (\ref{eq:N3S}) is consistently applied for spacetime with any dimension $D$.
As we will see below, $h_{\mu\nu}$ and $\phi$ decouple from $A_{\mu\nu\rho}$ and $D_{\mu}$ in the massless limit.
Thus $h_{\mu\nu}$ and $\phi$ might be regarded as the St\"{u}ckelberg-type fields needed for maintaining the gauge invariance of the massless part of the action after giving the mass for the fields $A_{\mu\nu\rho}$ and $D_{\mu}$.
However, unlike the St\"{u}ckelberg action for vector field or weak graviton field, the gauge transformation for one of the original fields $D_{\mu}$ is modified if we introduce the non-zero mass $m$ as can be seen in (\ref{eq:N3gaugeD}).
This fact is one reason why the action (\ref{eq:N3S}) has a rather complicated form.

The massless limit of the action $S_{\rm inv.}^{N=3,{\rm S}}$ is obtained by taking the limit $m\rightarrow 0$ of (\ref{eq:N3S}).
The resulting action $S_{{\rm inv.}, m\rightarrow 0}^{N=3,{\rm S}}$ is simplified after the following redefinition of $h_{\mu\nu}$ and $\phi$:
\begin{eqnarray}
\bar{h}_{\mu\nu} &=& h_{\mu\nu} +\frac{9}{3(D-2)} (\phi-h_{\rho}{}^{\rho}) \eta_{\mu\nu},
\\
\bar{\phi} &=& 6 \sqrt{\frac{D+1}{3(D-2)}}  (\phi-h_{\rho}{}^{\rho})
\end{eqnarray}
where $D$ is the spacetime dimension.
The action is divided into three parts 
\begin{equation} 
S_{{\rm inv.}, m\rightarrow 0}^{N=3,{\rm S}} = 
S^{N=3,{\rm S}}_{m\rightarrow 0; A_{\mu\nu\rho},D_\mu} +S^{N=3,{\rm S}}_{m\rightarrow 0; h_{\mu\nu}}
+S^{N=3,{\rm S}}_{m\rightarrow 0; \phi}.
\end{equation}
Also, the gauge transformations (\ref{eq:N3Str1})$\sim$(\ref{eq:N3Str4}) in the $m\rightarrow 0$ limit are divided into three independent parts.
The explicit form of each action with gauge transformations is given as follows.
The action for $A_{\mu\nu\rho}$ and $D_\mu$ is 
\begin{eqnarray}
S^{N=3,{\rm S}}_{m\rightarrow 0; A_{\mu\nu\rho},A_\mu} &=&
  \frac{3}{2}  \int\!\!{d^{D}x} 
\Big[ A_{\mu\nu\rho} \Box A^{\mu\nu\rho}
+3 \partial_\rho A_{\mu\nu}{}^{\rho}  \partial_\sigma A^{\mu\nu\sigma} 
+ 3 \partial^\mu A_{\mu\nu}{}^{\nu}  \partial^\sigma A_{\sigma\rho}{}^{\rho} 
+3 A_{\mu\nu}{}^{\nu} \Box A^{\mu}{}_{\rho}{}^{\rho} 
\nonumber\\
&&
\qquad 
-\, 4D_\mu \left(\partial^\mu \partial^\nu - \eta^{\mu\nu} \Box \right)  A_{\nu\rho}{}^{\rho}
- \,4 D^{\rho} \partial^\mu \partial^\nu  A_{\mu\nu\rho} 
+\, 2 \partial_\mu D^{\mu} \partial_\nu  D^{\nu} \Big] 
\label{eq:SN31m01}
\end{eqnarray}
which is invariant under the gauge transformations 
\begin{equation}
\delta A_{\mu\nu\rho} =\partial_{(\mu} \lambda_{\nu\rho)},
\qquad
\delta D_{\mu} = - \partial^\nu \lambda_{\mu\nu} .
\label{eq:gtrN31m01}
\end{equation}
We further perform the following field redefinitions
\begin{eqnarray}
\tilde{A}_{\mu\nu\rho} &=&
A_{\mu\nu\rho} +3e \eta_{(\mu\nu}A_{\rho)\sigma}{}^\sigma
+2e \eta_{(\mu\nu}D_{\rho)} 
,
\\
\tilde{D}_{\mu} &=& d A_{\mu\rho}{}^\rho  \frac{2e-1}{3}A_\mu.
\end{eqnarray}
If we choose the parameter $e= (-1\pm \sqrt{1+D})/D$, the action (\ref{eq:SN31m01}) and the gauge transformations (\ref{eq:gtrN31m01}) become the following form:
\begin{eqnarray}
S^{N=3,{\rm S}}_{m\rightarrow 0; A_{\mu\nu\rho},A_\mu} 
&=&
3\int \!\!d^Dx\Big[
\frac{1}{2}\tilde{A}_{\mu\nu\rho}\Box\tilde{A}^{\mu\nu\rho}
+\frac{3}{2} \partial^{\mu}\tilde{A}_{\mu\rho\sigma} \partial_{\nu}\tilde{A}^{\nu\rho\sigma} 
\nonumber\\
&&\qquad\qquad\qquad
+6 \partial_{\mu}\partial_{\nu}\tilde{A}^{\mu\nu\rho} \tilde{D}_{\rho}
-6 \tilde{D}_\mu\Box \tilde{D}^\mu + \tilde{D}^\mu \partial_\mu \partial_\nu \tilde{D}^\nu
\Big]
\end{eqnarray}
and
\begin{equation}
\delta \tilde{A}_{\mu\nu\rho}=\partial_{(\mu}\tilde{\lambda}_{\mu\rho)},
\qquad
\delta \tilde{D}_{\mu}=\frac{1}{3} \partial^{\nu}\tilde{\lambda}_{\mu\nu}.
\end{equation}
Note that this form of the action naturally coincides with the so-called `triplet' action\,\cite{Francia:2002pt, Sagnotti:2003qa} 
obtained from the tensionless limit of the first Regge trajectory of open string theory.
We will discuss the general tensionless limit of our theory in the next section~5.
The other actions are the quadratic Einstein action for the metric $\bar{g}_{\mu\nu} \sim \eta_{\mu\nu}+\bar{h}_{\mu\nu}$ and the action for scalar field $\phi$:
\begin{equation}
S^{N=3,{\rm S}}_{m\rightarrow 0; A_{\mu}}=
8 [\sqrt{-\bar{g}} \, \bar{R}] \Big|_{2} 
\end{equation}
and
\begin{equation}
S^{N=3,{\rm S}}_{m\rightarrow 0; \phi}=
-\frac{1}{2}\bar{\phi} \Box \bar{\phi}.
\end{equation}
The gauge transformations for $h_{\mu\nu}$ and $\phi$ are naturally given by
\begin{equation}
\delta \bar{A}_{\mu\nu} = 2 \partial_{(\mu} \tilde{\lambda}_{\nu)}
\end{equation}
and
\begin{equation}
\delta \bar{\phi} =  0.
\end{equation}

\subsubsection{Action for massive anti-symmetric tensor field $B_{[\mu\nu]}$ with
$C_{\mu}$}
The action $S_{\rm inv.}^{N=3,{\rm A}}$ for the anti-symmetric tensor field 
$B_{\mu\nu}$ and the vector field $C_\mu$ of (\ref{eq:N3Sredef5}) and (\ref{eq:N3Sredef6}) is calculated as
\begin{equation}
S_{\rm inv.}^{N=3,{\rm A}} =
\int{d^{D}x} \left[
-\frac{1}{24}H_{\mu\nu\rho}H^{\mu\nu\rho}
-\frac{1}{8} (mB_{\mu\nu} +2 \partial_{[\mu}C_{\nu]})
(mB^{\mu\nu} +2 \partial^{[\mu}C^{\nu]})
\right]
\end{equation}
where $m=\sqrt{2/\alpha'}$ and 
$H_{\mu\nu\rho}=\partial_{\mu}B_{\nu\rho}+
\partial_{\nu}B_{\rho\mu}+\partial_{\rho}B_{\mu\nu}$.
We see that the vector $C_\mu$ is the St\"{u}ckelberg field.
The gauge transformations for $B_{[\mu\nu]}$ and $C_\mu$ are given by 
\begin{equation}
\delta B_{\mu\nu} =\partial_\mu\bar{\lambda}_\nu - \partial_\nu\bar{\lambda}_\mu
,\qquad
\delta C_{\mu} =-m \bar{\lambda}_\mu +\partial_\mu \bar{\lambda}
\end{equation}
where $\bar{\lambda}_\mu$ and $\bar{\lambda}$ are arbitrary vector and scalar fields respectively.

In the massless limit $m\rightarrow 0$,
the action becomes 
\begin{equation}
S_{{\rm inv.},m\rightarrow 0}^{N=3,{\rm A}} =
\int{d^{D}x} \left[
-\frac{1}{24}H_{\mu\nu\rho}H^{\mu\nu\rho}
-\frac{1}{8} H_{\mu\nu}H^{\mu\nu}
\right]
\end{equation}
with the field strength $H_{\mu\nu}= \partial_\mu C_\nu- \partial_\nu C_\mu$ for $C_{\mu}$.
This is the sum of the actions for massless anti-symmetric field $B_{\mu\nu}$ and the 
massless vector field $C_{\mu}$.
As is expected, the two parts are separately invariant under the transformations
\begin{equation}
\delta B_{\mu\nu} =\partial_\mu\bar{\lambda}_\nu - \partial_\nu\bar{\lambda}_\mu
,\qquad
\delta B_{\mu} = \partial_\mu \bar{\lambda} .
\end{equation}

\subsubsection{Comment on gauge fixed actions}
We comment on the gauge fixed actions for $S^{N=3,{\rm S}}_{\rm inv.}$ and $S^{N=3,{\rm A}}_{\rm inv.}$.
For $N=3$, non-trivial
 fields in the (anti-)ghost and gauge fixing terms are provided by SU(1,1)-spin $>0$ part of 
$\phi^{(n)}$ ($n=0, \pm 1, \pm 2$) and $\beta^{(m)}$ ($m=0,1,2$).
Among them, a part of  $\phi^{(\pm 1)}$ and $\beta^{(1)}$ is used for the gauge fixed action for 
 $S^{N=3,{\rm S}}_{\rm inv.}$. 
The other fields, $\phi^{(\pm 2)}$ and SU(1,1)-spin $=1$ part of $\phi^{(0)}$ and 
a part of $\phi^{(\pm 1)}$ and $\beta^{(1)}$, are used for the gauge fixed action for $S^{N=3,{\rm A}}_{\rm inv.}$, 
which means that there appear (anti-)ghost for (anti-)ghost fields as well as (anti-)ghost fields in the latter action.

\section{Tensionless limit}
We consider the tensionless limit of the minimal gauge invariant action $S^{\rm min}_{\rm inv.}$ and the gauge fixed action $S^{\rm min}_{\{\alpha\}}$.
The limit can be consistently obtained by taking the $\alpha'\rightarrow \infty$ limit of the operators $\tilde{Q}$ and $P_{n}$ appearing in the actions.
In the limit, $\tilde{Q}$ becomes
\begin{equation}
\tilde{Q}= \sqrt{2\alpha'} \left( \tilde{Q}' +{\cal O}( 1/\alpha' )
 \right)
\end{equation}
where 
\begin{equation}
\tilde{Q}' =
p_\mu \sum_{n=1}^{\infty}
\left( \alpha_{-n}^\mu c_n +\alpha_{n}^\mu c_{-n} \right) 
\end{equation}
and thus $P_n$ becomes 
\begin{equation}
P_n' = -\frac{2}{p^2}  \tilde{Q}' \tilde{W}_-^{\langle n+1 \rangle} \tilde{Q}'
\end{equation}
since $L_0\,(=\alpha' p^2 +N-1)\rightarrow \alpha' p^2$ in the limit.
Thus, the gauge invariant action in the limit becomes 
\begin{equation}
S^{\rm min}_{{\rm inv.},\alpha'\rightarrow \infty} = 
-\frac{1}{2} \left\langle  \phi^{(0)}, c_0 \frac{p^2}{2} (1-P'_0)\phi^{(0)}\right\rangle
\label{eq:Sminainflim}
\end{equation}
where we have rescaled $\sqrt{2\alpha'}\phi^{(0)}\rightarrow \phi^{(0)}$.
This action is invariant under the gauge transformation 
\begin{equation}
\delta \phi^{(0)} = \tilde{Q}' \lambda^{(-1)}.
\label{eq:gaugetrainflim}
\end{equation}
Note that the action $S^{\rm min}_{{\rm inv.},\alpha'\rightarrow \infty}$ 
with the gauge invariance can be applied for any $D$ dimensional spacetime
since the relation $\tilde{Q}'{}^2= -\frac{1}{2} p^2 M$ holds for any $D$.
Note also that the above gauge invariant action (\ref{eq:Sminainflim}) can be rewritten as
\begin{equation}
S^{\rm min}_{{\rm inv.},\alpha'\rightarrow \infty} = 
-\frac{1}{2} \left\langle  \phi^{(0)}|_{s=0}, c_0 
\left(\frac{p^2}{2} + \tilde{Q}'M^- \tilde{Q}' \right)\phi^{(0)}|_{s=0}\right\rangle
\label{eq:Sminainflim_BO}
\end{equation}
where $\phi^{(0)}|_{s=0}$ is the SU(1,1)-spin $=0$ part of $\phi^{(0)}$.
This form of the action exactly coincides with the gauge invariant action 
proposed in refs.\cite{Bengtsson:1986ys, Ouvry:1986dv}. 

On the other hand, the gauge fixed action (\ref{eq:Sminord}) in the $\alpha'\rightarrow \infty$ limit becomes
\begin{eqnarray}
S_{\{\alpha\},\, \alpha'\rightarrow \infty}^{\rm min} 
&=& S^{\rm min}_{{\rm inv.},\alpha'\rightarrow \infty}  + S_{{\rm gh+gf},\{\alpha\}, \alpha'\rightarrow \infty} 
\nonumber\\
\!\!&=&\!\! 
-\frac{1}{2}\sum_{n=-\infty}^\infty \langle\phi^{(n+1)}, c_0\frac{p^2}{2}(1-P'_{-n-1})\phi^{(-n-1)}\rangle
+
\sum_{n=-\infty}^\infty \langle c_0 \beta^{(n+2)}, \tilde{W}^{\langle -n \rangle}_{-} \tilde{Q}'\phi^{(-n-1)} \rangle
\nonumber\\
&&\!\!
+\frac{1}{2}\sum_{n=-\infty}^\infty 
\,\,\sum_{k\in \{\frac{{\rm max}(|-n|,|n+2|) }{2} +\pmb{Z}_{\ge 0}\}} \!\!\alpha^k_{(-n,n+2)} 
\langle S_k \beta^{(-n)}, c_0 \tilde{W}_-^{\langle n+2 \rangle} S_k\beta^{(n+2)}\rangle
\label{eq:Smingfainf}
\end{eqnarray}
where we have rescaled $\sqrt{2\alpha'}\phi^{(n)}\rightarrow \phi^{(n)}$.
As for the gauge invariant action, this action is consistent for any spacetime dimension $D$.
This action is invariant under the BRST and the anti-BRST transformations 
\begin{eqnarray}
\delta_{\rm B} \phi^{(n)} &=& \eta \beta^{(n)} \qquad (n\ge 1),
\\
\delta_{\rm B} \phi^{(-n)} &=& \eta \left[
S_{n/2 }\tilde{Q}' \phi^{(-n-1)} + 
M W_{n+2} M^{n+1} \beta^{(-n)} 
\right]
\qquad (n\ge 0),
\\
\delta_{\rm B} \beta^{(\pm n)} &=& 0 \qquad (n\ge 0),
\end{eqnarray}
and
\begin{eqnarray}
\delta'_{\rm B} \phi^{(n)} &=& 
\eta' \left[
S_{n/2 } \tilde{Q}' M^nW_{n+1}\phi^{(n+1)} + 
M^{n+1} W_{n+2} \beta^{(n+2)} 
\right]
\qquad (n\ge 0),
\\
\delta'_{\rm B} \phi^{(-n)} &=& 
- \eta' W_{n} M^{n-1} \beta^{(-n+2)} 
\qquad (n\ge 1),
\\
\delta'_{\rm B} \beta^{(\pm n)} &=& 0 \qquad (n\ge 0)
\end{eqnarray}
which are given by the $\alpha'\rightarrow \infty$ limit of (\ref{eq:BRST1})$\sim$(\ref{eq:antiBRST3}).
Also, the propagator is consistently given by the  $\alpha'\rightarrow \infty$ limit of 
(\ref{eq:propDelta}) as
\begin{equation}
\Delta^{'\langle -n-1 \rangle} =
\frac{2}{p^2} \left[ 1-P'_{-n-1} 
-\sum_k \alpha^k_{(-n, n+2)} \frac{2}{p^2} \tilde{Q}' S_k \tilde{W}_-^{{\langle -n \rangle}}\tilde{Q}'
\right].
\end{equation}

These $S^{\rm min}_{{\rm inv.},\alpha'\rightarrow \infty}$ and $S_{\{\alpha\},\, \alpha'\rightarrow \infty}^{\rm min}$ are the sum of the actions for various massless fields of various integer spin or symmetry.
Each of these actions can be obtained by restricting $\phi^{(0)}$ (or $\phi^{(n)}$ for the gauge fixed action) to the set of fields which is decoupled from the other fields in the action.
For example, if we take
\begin{eqnarray}
\phi^{(0)}_{n, {\rm sym}} &=&
\int\!\!\frac{d^Dp}{(2\pi)^D}
\bigg[
\sqrt{\frac{2}{n!}}\,
\alpha_{-1}^{\mu_1} \alpha_{-1}^{\mu_2} \cdots \alpha_{-1}^{\mu_n} \ket{0, p; \downarrow}
A_{\mu_1\mu_2\cdots \mu_n}
\nonumber\\
&&\qquad - \sqrt{\frac{2n(n-1)}{(n-2)!}}\,
\alpha_{-1}^{\mu_1} \alpha_{-1}^{\mu_2} \cdots \alpha_{-1}^{\mu_{n-2}} b_{-1}c_{-1}\ket{0, p; \downarrow}
D_{\mu_1\mu_2\cdots \mu_{n-2}}\bigg]
\end{eqnarray}
for an arbitrary positive integer $n$ (including $n=1$ with $D=0$), 
the part of the action which contains $\phi^{(0)}_{n, {\rm sym}}$ is completely decoupled from other fields in the action $S^{\rm min}_{{\rm inv.},\alpha'\rightarrow \infty}$.
In fact, by substituting $\phi^{(0)}_{n, {\rm sym}}$ into (\ref{eq:Sminainflim}),  
we obtain 
\begin{eqnarray}
S^{n, {\rm sym}}_{{\rm inv.},\alpha'\rightarrow \infty}
\!\!&\!\!=\!\!& \!\!  \int \! d^D \!x \Big[
-\frac{1}{2} \partial_\nu A_{\mu_1\cdots \mu_n}  \partial^\nu A^{\mu_1\cdots \mu_n} 
+n(n-1)  \partial_\nu D_{\mu_1\cdots \mu_{n-2}}  \partial^\nu D^{\mu_1\cdots \mu_{n-2}}  
\nonumber\\
&&+\frac{n}{2} \partial_\mu A^{\mu \mu_1\cdots \mu_{n-1}} 
 \partial^\nu A_{\nu \mu_1\cdots \mu_{n-1}}  
+n(n-1) D_{\mu_1\cdots \mu_{n-2}} \partial_\mu \partial_\nu   
A^{\mu\nu\mu_1\cdots \mu_{n-2}}
\nonumber\\
&& + \frac{n(n-1)(n-2)}{2} \partial_\mu D^{\mu \mu_1\cdots \mu_{n-3}} 
 \partial^\nu D_{\nu \mu_1\cdots \mu_{n-3}}  
\Big].
\label{eq:Sinvnsym}
\end{eqnarray}
The gauge transformations for this action is obtained by substituting 
\begin{equation}
\lambda^{(-1)}_{n, {\rm sym}} = 
\int\!\!\frac{d^Dp}{(2\pi)^D}\alpha_{-1}^{\mu_1} \alpha_{-1}^{\mu_2} \cdots \alpha_{-1}^{\mu_{n-1}} b_{-1}
\ket{0, p; \downarrow} i \lambda_{\mu_1\mu_2\cdots \mu_{n-1} }
\end{equation}
into (\ref{eq:gaugetrainflim}) and the result in the $x$-representation is
\begin{equation}
\delta A_{\mu_1\cdots \mu_n} = \partial_{(\mu_1} \lambda_{\mu_2\cdots \mu_{n})},
\qquad
\delta D_{\mu_1\cdots \mu_{n-2}} = \frac{1}{n} \partial^{\mu} 
\lambda_{\mu \mu_1 \cdots \mu_{n-2}}.
\end{equation}
This action coincides with the `triplet' action investigated in refs.\cite{Francia:2002pt, Sagnotti:2003qa}.
Unlike the action given in refs.\cite{Francia:2002pt, Sagnotti:2003qa} which consists of the triplet of rank $n$, $n-1$ and $n-2$ symmetric tensor fields, 
our action (\ref{eq:Sinvnsym}) rather consists of a `doublet' of 
rank $n$ and $n-2$ symmetric tensor fields and does not contain the auxiliary rank $n-1$ symmetric tensor from the beginning. 
This is because all the auxiliary string fields has been eliminated from our minimal action in advance.

As for the (anti-)ghost and the gauge fixing terms for (\ref{eq:Sinvnsym}), we need the following set of string fields: 
\begin{eqnarray}
\phi^{(1)}_{n, {\rm sym}} &=&
\int\!\!\frac{d^Dp}{(2\pi)^D}
\sqrt{\!\frac{2}{(n-1)!}}\,
\alpha_{-1}^{\mu_1} \alpha_{-1}^{\mu_2} \cdots \alpha_{-1}^{\mu_{n-1}} c_{-1}
\ket{0, p; \downarrow} i \bar{\gamma}_{\mu_1\mu_2\cdots \mu_{n-1} },
\\
\phi^{(-1)}_{n, {\rm sym}} &=&
\int\!\!\frac{d^Dp}{(2\pi)^D}
\sqrt{\!\frac{2}{(n-1)!}}\,
\alpha_{-1}^{\mu_1} \alpha_{-1}^{\mu_2} \cdots \alpha_{-1}^{\mu_{n-1}} b_{-1}
\ket{0, p; \downarrow} {\gamma}_{\mu_1\mu_2\cdots \mu_{n-1} },
\\
\beta^{(1)}_{n, {\rm sym}} &=&
\int\!\!\frac{d^Dp}{(2\pi)^D}
\sqrt{\frac{2}{n!}}\,
\alpha_{-1}^{\mu_1} \alpha_{-1}^{\mu_2} \cdots \alpha_{-1}^{\mu_{n-1}} c_{-1}
\ket{0, p; \downarrow} i \beta_{\mu_1\mu_2\cdots \mu_{n-1} }.
\end{eqnarray}
Here $\gamma$ and $\bar{\gamma}$ are ghost and anti-ghost fields respectively, and 
$\beta$ is the Lagrange multiplier field needed for fixing the gauge symmetry.
Note that $\phi^{(\pm n)}$ for $n\ge 2$ do not couple to the above fields and 
(anti-)ghost for (anti-)ghost fields do not appear. 
Then the (anti-)ghost and gauge fixing terms are calculated by substituting 
$\phi^{(\pm 1)}_{n, {\rm sym}} $ and $\beta^{(1)}_{n, {\rm sym}} $ into 
\begin{eqnarray}
S_{{\rm gh+gf},\alpha, \alpha'\rightarrow \infty} 
&=&
- \Big\langle\phi^{(1)}_{n, {\rm sym}}, c_0\frac{p^2}{2}(1-P_{-1}')\phi^{(-1)}_{n, {\rm sym}}\Big\rangle
\nonumber\\
&&\qquad + \big\langle c_0 \beta^{(1)}_{n, {\rm sym}} , c_0 W_{1} \tilde{Q}'\phi^{(0)}_{n, {\rm sym}} \big\rangle
+\frac{1}{2}\alpha 
\big\langle \beta^{(1)}_{n, {\rm sym}} , c_0 {W}_{1} \beta^{(1)}_{n, {\rm sym}} \big\rangle
\end{eqnarray}
and the result is 
\begin{eqnarray}
S_{{\rm gh+gf},\{\alpha\}, \alpha'\rightarrow \infty}^{n, {\rm sym}}
&=&
\beta^{\mu_1\cdots\mu_{n-1}} \partial^\mu A_{\mu\mu_1\cdots\mu_{n-1}}
+(n-1) \partial_\mu \beta^{\mu\mu_1\cdots\mu_{n-2}} D_{\mu_1\cdots\mu_{n-2}}
\nonumber\\
&& +\frac{\alpha}{2n} \beta_{\mu_1\cdots\mu_{n-1}}\beta^{\mu_1\cdots\mu_{n-1}}
-i \partial_\mu\bar{\gamma}^{\mu_1\cdots\mu_{n-1}} 
\partial^\mu\gamma_{\mu_1\cdots\mu_{n-1}}.
\label{eq:Sghgfnsym}
\end{eqnarray}
Here, $\alpha$ is a real parameter.
The total gauge fixed action 
\begin{equation}
S_{\alpha,\, \alpha'\rightarrow \infty}^{n, {\rm sym}} 
= S^{n, {\rm sym}}_{{\rm inv.},\alpha'\rightarrow \infty}
  + S_{{\rm gh+gf},\alpha, \alpha'\rightarrow \infty}^{n, {\rm sym}}
\end{equation}
given by the sum of (\ref{eq:Sinvnsym}) and (\ref{eq:Sghgfnsym}) 
is the generalization of the gauge fixed action for vector field $A_\mu$.
In fact, for $n=1$, the action is reduced to the level $N=1$ action given in (\ref{eq:SminN1}).
Note also that the action $S_{\alpha,\, \alpha'\rightarrow \infty}^{n, {\rm sym}}$ 
for the particular value of $\alpha=0$ and $\alpha=1$ correspond to the generalization of
the Landau gauge and the Feynman gauge respectively.
The BRST and anti-BRST transformations for $S_{\alpha,\, \alpha'\rightarrow \infty}^{n, {\rm sym}}$ are respectively given by
\begin{eqnarray}
\delta_{\rm B} A_{\mu_1\mu_2\cdots \mu_{n} } &=& 
- i \eta \partial_{(\mu_1} \gamma_{\mu_2\cdots \mu_{n})} ,
\\
\delta_{\rm B} D_{\mu_1\mu_2\cdots \mu_{n-2}} &=& -i \eta \frac{1}{n}
\partial^{\mu} \gamma_{\mu\mu_1\cdots \mu_{n-2}},
\\
\delta_{\rm B} \bar{\gamma}_{\mu_1\mu_2\cdots \mu_{n-1}} &=& 
\frac{1}{n}\eta \beta_{\mu_1\mu_2\cdots \mu_{n-1}},
\\
\delta_{\rm B}{\gamma}_{\mu_1\mu_2\cdots \mu_{n-1}} &=& 0,
\\
 \delta_{\rm B} {\beta}_{\mu_1\mu_2\cdots \mu_{n-1}} &=& 0
\end{eqnarray}
and 
\begin{eqnarray}
\delta'_{\rm B} A_{\mu_1\mu_2\cdots \mu_{n} } &=& 
 \eta' \partial_{(\mu_1} \bar{\gamma}_{\mu_2\cdots \mu_{n})} ,
\\
\delta'_{\rm B} D_{\mu_1\mu_2\cdots \mu_{n-2}} &=& 
 \eta' \frac{1}{n}
\partial^{\mu} \bar{\gamma}_{\mu\mu_1\cdots \mu_{n-2}},
\\
\delta'_{\rm B} \bar{\gamma}_{\mu_1\mu_2\cdots \mu_{n-1}} &=& 
0,
\\
\delta'_{\rm B}{\gamma}_{\mu_1\mu_2\cdots \mu_{n-1}} &=& -i\frac{1}{n}\eta' \beta_{\mu_1\mu_2\cdots \mu_{n-1}},
\\
 \delta'_{\rm B} {\beta}_{\mu_1\mu_2\cdots \mu_{n-1}} &=& 0.
\end{eqnarray}
These transformations are reduced to (\ref{eq:A1BRST}) and  (\ref{eq:A1aBRST}) for $n=1$.

We can also extract gauge invariant action for rank $n$ antisymmetric tensor field 
$B_{\mu_1\mu_2\cdots\mu_n}(=B_{[\mu_1\mu_2\cdots \mu_n]})$ from $S^{\rm min}_{{\rm inv.},\alpha'\rightarrow \infty}$
by restricting the SU(1,1)-spin $=0$ part of $\phi^{(0)}$ as
\begin{equation}
\phi^{(0)}_{n, {\rm antisym}}|_{s=0} =
\int\!\!\frac{d^Dp}{(2\pi)^D}
2\left( \prod_{k=i}^n \ell_k \right)^{-\frac{1}{2}}
\alpha_{-\ell_1}^{[\mu_1} \alpha_{-\ell_2}^{\mu_2} \cdots \alpha_{-\ell_n}^{\mu_n]} \ket{0, p; \downarrow}
B_{\mu_1\mu_2\cdots \mu_n}
\end{equation}
where $\ell_k$ ($k=1,\cdots,n$) are arbitrary integers which differ from one another, {\it i.e.}, $\ell_k \ne \ell_j$ for $k\ne j$. 
Then the gauge invariant action is calculated by
substituting $\phi^{(0)}_{n, {\rm antisym}}|_{s=0}$ into (\ref{eq:Sminainflim}) 
and the result in the $x$-representation takes the following expected form 
\begin{equation}
S^{n, {\rm antisym}}_{{\rm inv.},\alpha'\rightarrow \infty}
=\int d^Dx \left[
-\frac{1}{n+1} H_{\mu_1\mu_2\cdots\mu_{n+1}}H^{\mu_1\mu_2\cdots\mu_{n+1}}
\right]
\end{equation}
where $H_{\mu_1\mu_2\cdots\mu_{n+1}}$ is the field strength for $B_{\mu_1\cdots\mu_{n}}$ 
given by
\begin{equation}
H_{\mu_1\mu_2\cdots\mu_{n+1}}
=(n+1) \partial_{[\mu_1}B_{\mu_2\cdots\mu_{n+1}]} .
\end{equation}
Gauge transformation for the $B_{\mu_1\cdots\mu_n}$ field is given by use of 
the SU(1,1)-spin $=\frac{1}{2}$ part of the corresponding $\lambda^{(-1)}$ string field 
\begin{equation}
\lambda^{(-1)}_{n, {\rm antisym}}|_{s=\frac{1}{2}} =
\!\int\!\!\frac{d^Dp}{(2\pi)^D}
\frac{2}{n}\left( \prod_{k=i}^n \ell_k \!\!\right)^{-\frac{1}{2}}\!
\sum_{k=1}^n
(-1)^{k-1}\alpha_{-\ell_1}^{[\mu_1}\cdots \widehat{{\alpha}_{-\ell_k}^{\mu_k}} \cdots \alpha_{-\ell_n}^{\mu_n]} b_{-\ell_k}\ket{0, p; \downarrow}
i\lambda_{\mu_1\mu_2\cdots \mu_{n-1}}
\end{equation}
where the oscillators on hat symbol are to be replaced by 1.
Explicitly, the gauge transformation 
\begin{equation}
\delta \phi^{(0)}_{n, {\rm antisym}}|_{s=0} =
S_{0}\tilde{Q}' \lambda^{(-1)}_{n, {\rm antisym}}|_{s=\frac{1}{2}} 
\end{equation}
is reduced again to the expected form  
\begin{equation}
\delta B_{\mu_1\cdots \mu_n} = \partial_{[\mu_1}\lambda_{\mu_2\cdots\mu_n]}.
\end{equation}
The gauge fixed action for general rank $n$ antisymmetric tensor field 
$B_{\mu_1\cdots \mu_n}$ is rather complicated because 
we need up to $n$-th rank of (anti-)ghost fields with the Lagrange multiplier fields given by $\beta^{(m)}$ in order to fix the gauge invariance completely.
More precisely, to fix the original gauge invariance,
we need SU(1,1)-spin $=\frac{1}{2}$ part of $\phi^{(\pm 1)}$ and $\beta^{(1)}$.
Then to fix the remaining gauge invariance, 
we next need SU(1,1)-spin $={1}$ part of $\phi^{(\pm 2)}$, $\phi^{(0)}$, $\beta^{(0)}$ and $\beta^{(2)}$.
In total, we need SU(1,1)-spin $=\frac{k}{2}$ part of $\phi^{(m)}\in {\cal F}^m$ and 
$\beta^{(m)}\in \tilde{\cal F}^m$ for $m=1,2,\cdots,n$.

Other than the totally symmetric or anti-symmetric tensor fields discussed above, 
the general action (\ref{eq:Smingfainf}) contains gauge invariant and gauge fixed actions for general massless fields of various mixed symmetry.

\section{Summary and discussions}
We have investigated the minimal gauge invariant action $S^{\rm min}_{\rm inv.}$ 
which is extracted from the quadratic part of the original covariant open bosonic string field theory
and have succeeded in constructing a family of the corresponding gauge fixed action 
$S^{\rm min}_{\{\alpha\}}$ which is parametrized by infinite number of real parameters $\alpha^k_{(-n,n+2)}$. 
The action $S^{\rm min}_{\{\alpha\}}$ is invariant under 
BRST and anti-BRST transformations and from which propagators are systematically obtained.
We have found that our result could be used as a powerful tool for 
constructing gauge invariant and gauge fixed actions
for any type of quantum fields as long as they are included in the spectrum of bosonic string theory.
These actions have sufficiently general and simple form.

As an example, we have investigated the actions for lower level ($N\le 3$) string fields.
We have found the complete gauge fixed action for weak massive graviton field $h_{\mu\nu}$ with $g_\mu$ and $\phi$ from $N=2$ part, and 
the gauge invariant action for massive 3rd rank symmetric tensor field $A_{\mu\nu\rho}$
with lower rank tensor fields and the antisymmetric tensor field $B_{[\mu\nu]}$ with a vector field from $N=3$ part.

We have also investigated the tensionless limit of the gauge invariant and gauge fixed actions, $S^{\rm min}_{{\rm inv.},\alpha'\rightarrow \infty}$ and $S^{\rm min}_{\{\alpha\},\alpha'\rightarrow \infty}$, which describe the actions for various massless fields.
Among them, in particular, we have identified the  
$n$-th rank totally symmetric tensor field part and anti-symmetric tensor field part. 
Other than these particular parts, 
it should be possible to identify and classify the actions for massless fields represented by tensor fields which have general mixed symmetry.

As for the massive fields, actions for general mixed symmetric tensor fields 
should also be extracted from $S^{\rm min}_{{\rm inv.}}$ and $S^{\rm min}_{\{\alpha\}}$.
In this case, the classification of the actions would be more complicated than the case of massless fields given by the tensionless limit.
For one thing, 
for higher level $N$, the number of fields increases explosively and the actions for the corresponding fields become complicated 
since the operator $\tilde{Q}$ appearing in the action mixes various types of tensor fields in general.
Also, to maintain the gauge invariance of higher rank tensor fields
appearing in the higher level string fields, we need a number of generally lower rank
St\"{u}ckelberg-like fields in addition, except for totally anti-symmetric field in which case we already know how to introduce the St\"{u}ckelberg-like fields.
Thus, gauge invariant actions for general mixed symmetric tensor fields 
should be formed of a number of types of fields.
It would be interesting if we could classify all the massive gauge invariant actions with respect to the symmetry and could find a rule 
how St\"{u}ckelberg-like fields appear for a given set of fields which are gauge invariant in the massless limit.

Our actions only contain bosonic fields in flat spacetime since we start from the bosonic string field theory.
To consider the extension of the actions to contain fermionic fields is one of the important future directions.
This would be accomplished by investigating the superstring field theory.
If we only aim at obtaining the quadratic actions for various fermionic fields at the moment, we may be avoid the many known difficulties concerning interaction part. 
Extension of our results to the curved spacetime, especially to AdS spacetime which is important in studying the theory in terms of AdS/CFT correspondence, 
is another direction.
We may be able to extend our minimal actions to some curved spacetimes only by extending the commutation relations for oscillators, though the extension of the original string field theory to curved spacetime is difficult.

\section*{Acknowledgements}
The author would like to thank M.~Kato for useful discussions and comments.
The author also thanks the Yukawa Institute for Theoretical Physics at Kyoto University. Communications during the YITP workshop on ``Field Theory and String Theory'' (YITP-W-12-05) were useful.
She is also obliged to Y.~Hikida for suggesting related references.

\appendix 
\def\thesection{Appendix~\Alph{section}}
\renewcommand{\theequation}{\Alph{section}.\arabic{equation}}
\def\thesection{Appendix~\Alph{section}}
\setcounter{equation}{0}
\setcounter{figure}{0}
\def\thesection{Appendix~\Alph{section}}
\section{Structure of open string Fock space and the projection operators}
\label{app1}
We summarize the detail of the action of open string field theory and the $a$-gauges.
For more details, see refs.\cite{Asano:2006hk, Asano:2008iu}.
We write the Fock space of string states as
\begin{equation}
{\cal F}=\bigoplus^{\infty}_{n=-\infty}\left(
{\cal F}^{n-1}+c_0 {\cal F}^{n-2}
\right)
\end{equation}
where ${\cal F}^n$ is spanned by the states of the form 
\begin{equation}
\ket{f^{(n)}}=
\alpha_{-n_1}^{\mu_1}\cdots  \alpha_{-n_i}^{\mu_i}
c_{-l_1}\cdots  c_{-l_{j}}\, b_{-m_1}\cdots  b_{-m_{k}} \,
\ket{0,p;\downarrow} 
\label{eq:A1}
\end{equation}
with 
$n_a>0$, $l_a>0$, $m_a>0$, and $j-k=n$. 
Note that the level $N$ of the state is given by
$N=\sum_i n_i +\sum_j l_j +\sum_k m_k $.
The state $\ket{0,p;\downarrow}=\ket{0,p}\otimes\ket{\!\downarrow}$ is annihilated by $\alpha^{\mu}_n$, $c_n$ and $b_n$ ($n>0$).
We set the ghost number of $\ket{\downarrow}\,(=c_1\ket{0})$ to be 1.
Then the states in ${\cal F}^n$ and $c_0{\cal F}^n$ have ghost number 
$n+1$ and $n+2$ respectively.
All states in the space ${\cal F}^n$ are classified by the SU(1,1)-spin $s\in \{ \frac{|n|}{2}, \frac{|n|}{2}+1, \frac{|n|}{2}+2, \cdots \}$.
String field $\phi^{(n)}$ is expanded by the Fock states as $\ket{f_i}\in {\cal F}^{n}$ as $\phi^{(n)}=\sum_i \ket{f_i}_n \phi_{f_i}^n$.

On the space ${\cal F}^n$, two types of projection operator are defined.
One is $P_{n}$ defined on ${\cal F}^{n}$ given by
\begin{equation}
P_n=-\frac{1}{L_0}  \tilde{Q} M^n W_{n+1} \tilde{Q},
\qquad
P_{-n}=-\frac{1}{L_0}  \tilde{Q} W_{n+1} \tilde{Q} M^{n}
\end{equation}
for $n\ge 0$, which can be written in together by using  the operator $\tilde{W}_-^{\langle n \rangle}$ given in (\ref{eq:deftildeW}) as
\begin{equation}
P_n=-\frac{1}{L_0}  \tilde{Q} \tilde{W}_-^{\langle n+1 \rangle}
\tilde{Q}.
\label{eq:defPn}
\end{equation}
Note that $P_{\pm n}$ is well-defined only for the states with 
$L_0(=\alpha'p^2+N- 1)\ne 0$\,\cite{Asano:2006hk}.
The other is the operator $S_k$ which extracts the SU(1,1)-spin $=k$ part from a given state $\ket{f}$.
For a state $\ket{f}^{\pm n}\in {\cal F}^{\pm n}$ ($n\ge 0$), $S_k$ is explicitly given by
\begin{eqnarray}
S_k &=& M^{k+ \frac{n}{2}} W_{2k} M^{k-\frac{n}{2}} - M^{k+1+ \frac{n}{2}} W_{2k+2} M^{k+1-\frac{n}{2}} 
\qquad (\mbox{on ${\cal F}^n$})
\\
S_k &=& 
M^{k- \frac{n}{2}} W_{2k} M^{k+\frac{n}{2}} 
- M^{k+1- \frac{n}{2}} W_{2k+2} M^{k+1+\frac{n}{2}} 
\qquad (\mbox{on ${\cal F}^{-n}$})
\end{eqnarray} 
for $k \in \{ \frac{n}{2} +\pmb{Z}_{\ge 0}  \}$ and $S_k=0$ for otherwise.
From the properties  $ M^n W_n \ket{f^{(n)}} =\ket{f^{(n)}}$ and $ W_n M^n \ket{f^{(-n)}} =\ket{f^{(-n)}} $, we can easily verify the relation $S_k{}^2=S_k$.
Also, the following relations hold:
\begin{equation}
S_k S_{k'} =\delta_{kk'} S_k ,\qquad
[S_k, M]=0,\qquad [S_k, W_{\pm n}]=0 \mbox{~~(on ${\cal F}^{\pm n}$)}.
\end{equation}
The inner product of two different spin states, e.g.,
$S_k\ket{f}\in {\cal F}^{n}$ and $S_k\ket{g}\in {\cal F}^{-n}$, is always vanishes:  
$ \langle S_{k}\ket{f}  , c_0 S_{k'} \ket{g}  \rangle =0$  for $k\ne k'$.

\section{Some useful formulas}
\setcounter{equation}{0}
We give some useful formulas for explicitly calculating the 
inner products of string states appearing in the 
string field theory actions.
Inner product of string fields $A$ and $B$ are given by 
\begin{equation}
\langle A, B\rangle = \langle {\rm bpz}(A) \ket{B}
\end{equation}
where bpz$(A)$ denotes the BPZ conjugation defined by the following relations:
\begin{equation} 
{\rm bpz}(\ket{0,p}) = \bra{0,-p},
\quad
{\rm bpz}(\ket{0}) = \bra{0},
\end{equation}
\begin{equation} 
{\rm bpz}(b_{-n}) = (-1)^n b_{n},\quad
{\rm bpz}(c_{-n}) = (-1)^{n-1} c_{n},\quad
{\rm bpz}(\alpha^{\mu}_{-n}) = (-1)^{n-1} \alpha^{\mu}_{n},
\end{equation}
\begin{equation} 
{\rm bpz}\,(\alpha \beta) = (-1)^{|\alpha| |\beta|}  {\rm bpz}(\beta) \, {\rm bpz}(\alpha) 
\end{equation}
where $|\alpha|$ and $|\beta|$ are the Grassmann parity of $\alpha$ and $\beta$ respectively.
Also, the relations 
\begin{equation}
{\rm bpz}(\tilde{Q}) = -\tilde{Q},\qquad
{\rm bpz}(M^n) = (-1)^n M^n,\qquad
{\rm bpz}(W_n) = (-1)^n W_n
\end{equation}
are useful.
The normalization of the inner product is 
\begin{equation}
\bra{0, p;\downarrow} c_0 \ket{0,p';\downarrow }=(2\pi)^{26} \delta^{26}(p-p'). 
\end{equation}
The (anti-)commutation relations among $\alpha_{n}^\mu$, $c_{n}$ and $b_{n}$ are 
\begin{equation}
[\alpha_{m}^\mu, \alpha_{n}^\nu ] =m\eta^{\mu\nu}\delta_{m+n,0},
\qquad \{b_{m},c_n \}= \delta_{m+n,0},
\quad \{b_m,b_n\}=\{c_m,c_n\}=0.
\end{equation}
As for the BRST operator $Q=\tilde{Q}+c_0 L_0+b_0 M$,
$\tilde{Q}$ is explicitly given by 
\begin{equation}
\tilde{Q}= \sum_{n\ne 0} c_{-n} L_n^{({\rm m})}  
-\frac{1}{2} \sum_{\parbox{12mm}{\tiny\rule{5pt}{0pt}$mn\ne 0$\\$m+n\ne0$}} (m-n)\,:c_{-m} c_{-n} b_{n+m}: 
\end{equation}
where $L_n^{({\rm m})}$ is the matter part of total $L_n= L_n^{({\rm m})} + L_n^{({\rm g})} $ and is given by
\begin{equation}
L_n^{({\rm m})} = \frac{1}{2} \sum_{n=-\infty}^{\infty} : \alpha_{m-n}^\mu \alpha_{\mu, n}:,
\qquad
L_n^{({\rm g})} = \frac{1}{2} \sum_{n=-\infty}^{\infty} (m-n): b_{m+n} c_{-n}: -\delta_{m,0} .
\end{equation}


\section{Another anti-BRST transformation $\tilde{\delta}'_{\rm B}$ commuting with $\delta_{\rm B}$} 
\setcounter{equation}{0}
In section~\ref{sec:BaB}, we gave a definition of BRST and anti-BRST transformations for the gauge fixed action $S^{\rm min}_{\{\alpha\}}$.
Here we give another definition of anti-BRST transformation $\tilde{\delta}_{\rm B}'$
which commutes with the BRST transformation $\delta_{\rm B}$.

First, we assume that the transformation $\tilde{\delta}_{\rm B}'$ to be
\begin{eqnarray}
\tilde{\delta}'_{\rm B}  \phi ^{(n)} &=&
\eta' \!\!\!\!\!\! \sum_{k\in \{\frac{n}{2} +\pmb{Z}_{\ge 0}\}}\!\!\! \!\!
A_{k}^n
S_k\tilde{Q} M^n W_{n-1} S_{k+\frac{1}{2}} \phi^{(n+1)}
\nonumber\\
&& \qquad \qquad \qquad \qquad
+ \,
\eta' \!\!\!\!\!\! \sum_{k\in \{\frac{n+2}{2} +\pmb{Z}_{\ge 0}\}}\!\!\! \!\!
B_{k}^n
S_k M^{n+1} W_{n+2}  \beta^{(n+2)}
\\
\tilde{\delta}'_{\rm B}  \phi ^{(-n-1)}  &=&
\eta' \!\!\!\!\!\! \sum_{k\in \{\frac{n+2}{2} +\pmb{Z}_{\ge 0}\}} \!\!\!\! \!\!
S_{k-\frac{1}{2}}\!
\left[
A_{k-\frac{1}{2}}^{-n-1}
 \tilde{Q} W_{n+2} M^{n+1} S_{k} \phi^{(-n)}
 \right.
\nonumber\\
&& \qquad \qquad\qquad\qquad \qquad 
\left.
+ 
B_{k-\frac{1}{2}}^{-n-1}
W_{n+1}M^n  \beta^{(-n+1)}
\right]
\\
\tilde{\delta}'_{\rm B}  \beta^{(n+1)}  &=&
\eta' \!\!\!\!\! \sum_{k\in \{\frac{n+2}{2} +\pmb{Z}_{\ge 0}\}}\!\!\! \!\!
C_{k-\frac{1}{2}}^{-n-1}
S_{k-\frac{1}{2}} \tilde{Q} M^{n+1} W_{n+2}  S_k \beta^{(n+2)}
\\
\tilde{\delta}'_{\rm B}  \beta^{(-n)}  \!\!&=&\!
\eta' \!\!\!\!\! \sum_{k\in \{\frac{n+2}{2} +\pmb{Z}_{\ge 0}\}}\!\!\! \!\!
C_{k}^{-n}
S_{k} \tilde{Q} W_{n+1} M^{n}  S_{k+\frac{n}{2}} \beta^{(-n+1)}
\end{eqnarray}
where $n\ge 0$
and $A^n_k$, $B^n_k$ and $C^n_k$ are yet undetermined parameters.
Then, by imposing the condition 
\begin{equation}
\tilde{\delta}'_{\rm B} S^{\rm min}_{\{\alpha\}} =0, \quad
\tilde{\delta}_{\rm B}'{}^{2}=0, \quad 
[\delta_{\rm B},\, \tilde{\delta}'_{\rm B} ]=0,
\end{equation}
the coefficients $A_k^n$ and $C_{k}^n$ are written with respect to $B_k^n$ as 
\begin{eqnarray}
&&
C^{\pm n}_k = A^{\pm n}_k =-\frac{B^{-n}_k+B^{n-1}_{k+\frac{1}{2}} }{\alpha^k_{(n,-n+2)}}
\qquad \left(n\ge 0, \; k\ge \frac{n+2}{2} \right) ,
\label{eq:c6}
\\
&&
C^{n}_{\frac{n}{2}} = A^{\pm n}_{\frac{n}{2}} = -B^{-n-1}_{\frac{n+1}{2}}
\qquad\qquad \left(n\ge 0 \right).
\end{eqnarray}
Furthermore, the relation   
\begin{equation}
B^0_k = B^{-1}_{k-\frac{1}{2}}
\label{eq:B0}
\end{equation}
holds for $k\ge 1$.
Since the overall factor of $\tilde{\delta}'_{\rm B}$ can be determined freely, we choose 
$A^{0}_{0}$=1. Then from the above conditions (\ref{eq:c6})$\sim$(\ref{eq:B0}), all the coefficients for $k= \frac{1}{2}$ and are determined as
\begin{equation}
C^{0}_0 = - B^{-1}_{\frac{1}{2}} = -B^{0}_1 =1, \qquad
C^{\pm 1}_{\frac{1}{2}} = A^{\pm 1}_{\frac{1}{2}} = \frac{2}{\alpha_{(1,1)}^{\frac{1}{2}}},
\end{equation}
and also all the $B^n_{k}$ for $k=1$ are determined as
\begin{equation}
B^0_1 =-1, \qquad
B^{-2}_{1} = \frac{2}{\alpha_{(1,1)}^{\frac{1}{2}}}.
\end{equation}
If all the $B_{l}^n$'s for $l<k$ are known, $B_l^n$ for $l=k$ must satisfy the following 
set of independent equations
\begin{eqnarray}
B^{n}_k B^{-n}_k &=& B^{n-1}_{k-\frac{1}{2}} B^{-n-1}_{k-\frac{1}{2}}
\qquad (k\ge 3/2,\; n>0),
\label{eq:B1}
\\
\alpha^{k-\frac{1}{2}}_{(-n+1, n+1)}
( B^{-n+1}_{k-\frac{1}{2}} + B^{n-2}_{k}   )  
&=&  \alpha^{k-\frac{1}{2}}_{(-n+3, n-1)}
( B^{n-1}_{k-\frac{1}{2}} + B^{-n}_{k} )
\qquad (k\ge 2,\; n>1)
\label{eq:B2}
\end{eqnarray}
for $n=2k-2, 2k-4, 2k-6\cdots$, 
and 
\begin{equation}
\alpha^{k-\frac{1}{2}}_{(2k-1, -2k+3)}  B^{-2k}_k 
= B^{-2k+1}_{k-\frac{1}{2}} + B^{2k-2}_{k} \qquad (k \ge 3/2)
\label{eq:B3}.
\end{equation}
The total number of the above relations (\ref{eq:B1}), (\ref{eq:B2}) and (\ref{eq:B3})
are $2[k]-1$ where $[k]$ is the integer part of $k$.
The degrees of freedom of $B_{k}^{n}$ for a fixed $k$ is $2[k]$.
Thus,  
to determine all the $B_{k-\frac{1}{2}}^{n}$ and $B_{k}^n$ for a given integer $k\, (\ge 2)$, 
which have totally $4k-2$ degrees of freedom,
there are $(2k-3)+(2k-1)+1$ conditions including (\ref{eq:B0}).   
In fact, if we determine a value of $B^0_k = B^{-1}_{k-\frac{1}{2}}$ (e.g., to be $-1$),  
all the coefficients $B_{k-\frac{1}{2}}^{n}$ and $B_{k}^n$ are determined and we have a consistent anti-BRST transformation $\tilde{\delta}'_{\rm B}$ which commutes with the BRST transformation $\delta_{\rm B}$ given in the text in section~\ref{sec:BaB}.



\begin{thebibliography}{999}

\bibitem{Sorokinetal}
  D.~Sorokin,
  ``Introduction to the classical theory of higher spins,''
  AIP Conf.\ Proc.\  {\bf 767} (2005) 172
  [hep-th/0405069];
  N.~Bouatta, G.~Compere and A.~Sagnotti,
  ``An Introduction to free higher-spin fields,''
  hep-th/0409068;
  A.~Sagnotti, E.~Sezgin and P.~Sundell,
  ``On higher spins with a strong Sp(2,R) condition,''
  hep-th/0501156;
  X.~Bekaert, S.~Cnockaert, C.~Iazeolla and M.~A.~Vasiliev,
  ``Nonlinear higher spin theories in various dimensions,''
  hep-th/0503128.
  A.~Campoleoni,
  ``Metric-like Lagrangian Formulations for Higher-Spin Fields of Mixed Symmetry,''
  Riv.\ Nuovo Cim.\  {\bf 33} (2010) 123
  [arXiv:0910.3155 [hep-th]];
  X.~Bekaert, N.~Boulanger and P.~Sundell,
  ``How higher-spin gravity surpasses the spin two barrier: no-go theorems versus yes-go examples,''
  arXiv:1007.0435 [hep-th];
  A.~Fotopoulos and M.~Tsulaia,
  ``Gauge Invariant Lagrangians for Free and Interacting Higher Spin Fields. A Review of the BRST formulation,''
  Int.\ J.\ Mod.\ Phys.\ A {\bf 24} (2009) 1
  [arXiv:0805.1346 [hep-th]].

\bibitem{Bengtsson:1986ys}
  A.~K.~H.~Bengtsson,
  ``A Unified Action For Higher Spin Gauge Bosons From Covariant String Theory,''
  Phys.\ Lett.\ B {\bf 182} (1986) 321.
  
\bibitem{Ouvry:1986dv}
  S.~Ouvry and J.~Stern,
  ``Gauge Fields Of Any Spin And Symmetry,''
  Phys.\ Lett.\ B {\bf 177} (1986) 335.
  
\bibitem{Francia:2002pt}
  D.~Francia and A.~Sagnotti,
  ``On the geometry of higher spin gauge fields,''
  Class.\ Quant.\ Grav.\  {\bf 20} (2003) S473
  [hep-th/0212185].
  
\bibitem{Sagnotti:2003qa}
  A.~Sagnotti and M.~Tsulaia,
  ``On higher spins and the tensionless limit of string theory,''
  Nucl.\ Phys.\ B {\bf 682} (2004) 83
  [hep-th/0311257].


  
\bibitem{Francia:2006hp}
  D.~Francia and A.~Sagnotti,
  ``Higher-spin geometry and string theory,''
  J.\ Phys.\ Conf.\ Ser.\  {\bf 33} (2006) 57
  [hep-th/0601199].

\bibitem{Sagnotti:2011qp}
  A.~Sagnotti,
  ``Notes on Strings and Higher Spins,''
  arXiv:1112.4285 [hep-th].

\bibitem{Siegel:1985tw}
  W.~Siegel and B.~Zwiebach,
  ``Gauge String Fields,''
  Nucl.\ Phys.\ B {\bf 263} (1986) 105.

\bibitem{Banks:1985ff}
T.~Banks and M.~E.~Peskin,
``Gauge invariance of string fields,''
Nucl.\ Phys.\ B {\bf 264} (1986) 513.

\bibitem{Itoh:1985bb}
  K.~Itoh, T.~Kugo, H.~Kunitomo and H.~Ooguri,
  ``Gauge Invariant Local Action of String Field from BRS Formalism,''
  Prog.\ Theor.\ Phys.\  {\bf 75} (1986) 162.

\bibitem{Witten:1985cc}
  E.~Witten,
  ``Non-commutative geometry and string field theory,''
  Nucl.\ Phys.\ B {\bf 268} (1986) 253.

\bibitem{Sen:1999nx}
  A.~Sen and B.~Zwiebach,
  ``Tachyon condensation in string field theory,''
  JHEP {\bf 0003} (2000) 002
  [hep-th/9912249].
  
\bibitem{Schnabl:2005gv}
  M.~Schnabl,
  ``Analytic solution for tachyon condensation in open string field theory,''
  Adv.\ Theor.\ Math.\ Phys.\  {\bf 10} (2006) 433
  [hep-th/0511286].

\bibitem{Siegel:1984wx}
  W.~Siegel,
  ``Covariantly Second Quantized String. 2.,''
  Phys.\ Lett.\ B {\bf 149} (1984) 157
   [Phys.\ Lett.\ B {\bf 151} (1985) 391].

\bibitem{Bochicchio:1986bd}
  M.~Bochicchio,
  ``String Field Theory In The Siegel Gauge,''
  Phys.\ Lett.\ B {\bf 188} (1987) 330.
  
\bibitem{Bochicchio:1986zj}
  M.~Bochicchio,
  ``Gauge Fixing For The Field Theory Of The Bosonic String,''
  Phys.\ Lett.\ B {\bf 193} (1987) 31.
  
\bibitem{Thorn:1986qj}
  C.~B.~Thorn,
  ``Perturbation Theory for Quantized String Fields,''
  Nucl.\ Phys.\ B {\bf 287} (1987) 61.

\bibitem{Asano:2006hk}
  M.~Asano and M.~Kato,
  ``New covariant gauges in string field theory,''
  Prog.\ Theor.\ Phys.\  {\bf 117} (2007) 569
  [arXiv:hep-th/0611189];


\bibitem{Asano:2008iu}
  M.~Asano and M.~Kato,
  ``General Linear Gauges and Amplitudes in Open String Field Theory,''
  Nucl.\ Phys.\  B {\bf 807} (2009) 348
  [arXiv:0807.5010 [hep-th]].

\bibitem{Kiermaier:2007jg}
  M.~Kiermaier, A.~Sen and B.~Zwiebach,
  ``Linear b-Gauges for Open String Fields,''
  JHEP {\bf 0803} (2008) 050
  [arXiv:0712.0627 [hep-th]].

\bibitem{Siegel:1984xd}
  W.~Siegel,
  ``Covariantly Second Quantized String. 3,''
  Phys.\ Lett.\  B {\bf 149} (1984) 162
  [Phys.\ Lett.\  {\bf 151B} (1985) 396].

\bibitem{Zwiebach:1992ie}
  B.~Zwiebach,
  ``Closed string field theory: Quantum action and the B-V master equation,''
  Nucl.\ Phys.\  {\bf B390 } (1993)  33-152.
  [hep-th/9206084].


\bibitem{Asano:2012sk}
  M.~Asano and M.~Kato,
  ``Closed string field theory in a-gauge,''
  arXiv:1206.3901 [hep-th].

\bibitem{Preitschopf:fc}
C.~R.~Preitschopf, C.~B.~Thorn and S.~A.~Yost,
``Superstring field theory,''
Nucl.\ Phys.\ B {\bf 337}, 363 (1990).

\bibitem{Arefeva:1989cp}
I.~Y.~Arefeva, P.~B.~Medvedev and A.~P.~Zubarev,
``New representation for string field
solves the consistence problem for open superstring field,''
Nucl.\ Phys.\ B {\bf 341}, 464 (1990).

\bibitem{Kroyter:2012niTorii:2012nj}
  M.~Kroyter, Y.~Okawa, M.~Schnabl, S.~Torii and B.~Zwiebach,
  ``Open superstring field theory I: gauge fixing, ghost structure, and propagator,''
  JHEP {\bf 1203} (2012) 030
  [arXiv:1201.1761 [hep-th]];
  S.~Torii,
  ``Validity of Gauge-Fixing Conditions and the Structure of Propagators in Open Superstring Field Theory,''
  JHEP {\bf 1204} (2012) 050
  [arXiv:1201.1762 [hep-th]].

\bibitem{Kroyter:2009rn}
 M.~Kroyter,
 ``Superstring field theory in the democratic picture,''
arXiv:0911.2962.

\bibitem{Kohriki:2011zzKohriki:2011pp} 
  M.~Kohriki, I.~Kishimoto, T.~Kugo, H.~Kunitomo and M.~Murata,
  ``Gauge-fixing problem in modified cubic superstring field theory,''
  Prog.\ Theor.\ Phys.\ Suppl.\  {\bf 188}, 263 (2011);
  M.~Kohriki, T.~Kugo and H.~Kunitomo,
  ``Gauge fixing of modified cubic open superstring field theory,''
  arXiv:1111.4912.

\bibitem{Kroyter:2009zi} 
  M.~Kroyter,
  ``Superstring field theory equivalence: Ramond sector,''
  JHEP {\bf 0910}, 044 (2009)
  [arXiv:0905.1168].

\bibitem{Berkovits:2001im}
  N.~Berkovits,
	``The Ramond sector of open superstring field theory,''
  JHEP {\bf 0111}, 047 (2001) [hep-th/0109100].

\bibitem{Michishita:2004by}
  Y.~Michishita,
	``A covariant action with a constraint and Feynman rules
    for fermions in open superstring field theory,''
    JHEP {\bf 0501}, 012 (2005) [hep-th/0412215].

\bibitem{Fierz:1939ix}
  M.~Fierz and W.~Pauli,
  ``On relativistic wave equations for particles of arbitrary spin in an electromagnetic field,''
  Proc.\ Roy.\ Soc.\ Lond.\ A {\bf 173} (1939) 211.

\bibitem{Hinterbichler:2011tt}
  K.~Hinterbichler,
  ``Theoretical Aspects of Massive Gravity,''
  Rev.\ Mod.\ Phys.\  {\bf 84} (2012) 671
  [arXiv:1105.3735 [hep-th]].

\bibitem{Kato:1983im}
M.~Kato and K.~Ogawa,
``Covariant quantization of string based on BRS invariance,''
Nucl.\ Phys.\ B {\bf 212}, 443 (1983).


\bibitem{Taylor:2003gn}
  W.~Taylor and B.~Zwiebach,
  ``D-branes, tachyons, and string field theory,''
  hep-th/0311017.
  


\end{thebibliography}
\end{document}